\documentclass[a4paper,fleqn,usenatbib]{mnras}

\usepackage[T1]{fontenc}
\usepackage{ae,aecompl}

\usepackage{graphicx}	
\usepackage{amsmath}	
\usepackage{amssymb}	

\usepackage{txfonts}

\title[The IMF as a function of supersonic turbulence]{The IMF as a function of supersonic turbulence}
\author[C. Bertelli Motta, P. C. Clark, S. C. O. Glover, R. S. Klessen, A. Pasquali]{C. Bertelli Motta$^{1,2}$\thanks{E-mail:
cbertelli@ari.uni-heidelberg.de}, P. C. Clark$^{3,2}$, S. C. O. Glover$^{2}$, R. S. Klessen$^{2}$, A. Pasquali $^{1}$\\
$^{1}$Astronomisches Rechen-Institut, Zentrum f\"ur Astronomie der Universit\"at Heidelberg, M\"onchhofstr. 12-14, 69120 Heidelberg, Germany\\
$^{2}$Institute of Theoretical Astrophysics, Zentrum f\"ur Astronomie der Universit\"at Heidelberg, Albert-\"Uberle-Str. 2, 69120 Heidelberg, Germany\\
$^{3}$School of Physics and Astronomy, Cardiff University, The Parade, Cardiff CF24 3AA, UK
}
\begin{document}

\date{Accepted 2016 August 01. Received 2016 July 20; in original form 2015 September 07}

\pagerange{\pageref{firstpage}--\pageref{lastpage}} \pubyear{2015}

\maketitle

\label{firstpage}

\begin{abstract}
Recent studies seem to suggest that the stellar initial mass function (IMF) in early-type galaxies might be different from a classical Kroupa or Chabrier IMF, i.e. contain a larger fraction of the total mass in low-mass stars. From a theoretical point of view, supersonic turbulence has been the subject of interest in many analytical theories proposing a strong correlation with the characteristic mass of the core mass function (CMF) in star forming regions, and as a consequence with the stellar IMF. Performing two suites of smoothed particles hydrodynamics (SPH) simulations with different mass resolutions, we aim at testing the effects of variations in the turbulent properties of a dense, star forming molecular cloud on the shape of the system mass function in different density regimes.
While analytical theories predict a shift of the peak of the CMF towards lower masses with increasing velocity dispersion of the cloud, we observe in the low-density regime the opposite trend, with high Mach numbers giving rise to a top-heavy mass distribution. For the high-density regime we do not find any trend correlating the Mach number with the characteristic mass of the resulting IMF, implying that the dynamics of protostellar accretion discs and fragmentation on small scales is not strongly affected by turbulence driven at the scale of the cloud.
Furthermore, we suggest that a significant fraction of dense cores are disrupted by turbulence before stars can be formed in their interior through gravitational collapse. Although this particular study has limitations in its numerical resolution, we suggest that our results, along with those from other studies, cast doubt on the turbulent fragmentation models on the IMF that simply map the CMF to the IMF.
\end{abstract}


\section{Introduction}
The stellar initial mass function (IMF), i.e. the mass distribution of stars belonging to the same population at the time of their birth, is a key parameter in many fields of astrophysics. Beside being central to our understanding of the star formation processes, it is used as a tool to understand galactic evolution and cosmology, e.g. in the determination of the mass of distant galaxies whose stellar populations cannot be resolved, for galaxy evolution models, and for the search of planet-hosting stars.

\subsection{Observations of the IMF}
The first efforts to categorise the initial mass distribution of field stars in the solar neighbourhood was made by \citet{salpeter1955} by means of a single-slope power-law function and later by \citet{millerscalo1979} who introduced the idea of a lognormal distribution at low masses. Thereafter many more models have been developed, two of the most famous of which are the segmented power laws described by \citet{kroupa1993} and thereafter in \citet{kroupa2001}, later revisited in \citet{thies2007} and the combination of a lognormal and a power law function described by \citet{kroupa1990} and \citet{chabrier2003}. 

For many years the IMF has been thought to be universal and equal to the one observed in the Milky Way disc, not least due to the impossibility of a determination of the IMF by direct counting of the stars in more distant galaxies, where only very massive, bright stars, if any, can be resolved (see e.g. \citealt{hunter1996}, \citealt{malumuth1996}).  Recently, new techniques used to determine the mass of distant early-type galaxies (ETG) have shed new light on this topic. As \citet{vandokkum2010} pointed out, the presence of strong gravity-sensitive spectroscopic features typical of dwarf stars in the spectra of ETGs is not consistent with the assumption of a classical IMF, but points towards a bottom-heavy IMF, characterised by a higher fraction of mass included into low-mass stars than observed in the solar vicinity. This picture has been confirmed also by later works, e.g. \citet{vandokkum2011}, \citet{ferreras2013} and \citet{labarbera2013}.

Based on a different approach, \citet{cappellari2012} also questioned the validity of the usual assumption made about the universality of the IMF. With a dynamical study, they showed that the mass-to-light ratio of ETGs is higher than predicted by a Kroupa or Chabrier IMF. This indicates a larger amount of mass included into faint objects than expected assuming an underlying classical IMF, a picture that would be consistent with both a bottom- or a top-heavy IMF. In fact, such results could be explained by the presence of a large number of either low-mass, faint stars or compact remnants of high-mass, short-lived stars. 
Actually, neither scenario explains satisfactorily all of the observational evidence. A time-independent, bottom-heavy IMF cannot explain the amount of $\alpha$-elements measured in ETGs, which can only be produced by high-mass stars releasing their synthesized elements into the interstellar medium at the end of their life, while a top-heavy IMF is not consistent with the spectral features typical of dwarf stars observed by \citet{vandokkum2010}. 
For this reason, \citet{weidner2013} suggested the possibility of a time-dependent IMF which, starting as a top-heavy distribution at high redshift, turns into a bottom-heavy one at later times.

Despite the many efforts made in this sense, the universality of the IMF and the implied existence of fine-tuned processes leading to star formation remains a highly debated topic. In fact, in addition to the studies listed above, other works e.g. based on gravitational lensing effects \citep{smith2013} and on X-ray emission from low-mass X-ray binaries \citep{peacock2014} claim that the IMF of ETGs is consistent with an invariant IMF. For further discussion, see e.g. \citet{bastian2010}, \citet{offner2014}, \citet{krumholz2014} or \citet{klessenglover2014}.

\subsection{Theoretical IMF models}
\label{sec:introth}
From a theoretical point of view this discussion translates into the question about the physical processes capable of altering the shape of the IMF. What are they? How do they affect star formation? In the present work we focus on the effects of supersonic turbulence on the mass distribution of newly formed stars. Molecular clouds are known to be characterised by supersonic motions of the gas, which can be inferred from the broadening of spectral lines, inconsistent with the temperatures typical of such regions (see e.g. \citealt{maclowklessen2004} or \citealt{mckee2007}).

Several analytical theories which describe the effects of supersonic turbulence on the mass function of dense prestellar cores (the core mass function, or CMF) have been developed (for a comprehensive overview see e.g. \citealt{offner2014}, \citealt{krumholz2014} or \citealt{klessenglover2014}). If one assumes a simple mapping from the core mass function to the IMF, as suggested e.g.\ by the similar shape of the two functions \citep[see e.g.][]{alves2007,andre10}, then these theories also yield predictions for the IMF. For example, \citet{padoan2002} assume star formation to take place in the dense sheets and filaments formed by interacting shocks. They assume that the gas is isothermal and that the turbulence is supersonic, with a power spectrum $E(k)\propto k^{-\beta}$ with $\beta=2$. In this case, they argue that the mass spectrum of the collapsing dense cores for which the gravitational energy exceeds the thermal energy is given by the distribution:
\begin{equation}
 N(m)\rmn{d}\log m\propto m^{-3/(4-\beta)} \left[\int_0^m p(m_{\rm J}) \rmn{d}m_{\rm J}\right]\rmn{d}\log m
\end{equation}
where $p(m_{\rm J})$ is the probability distribution for the local Jeans mass. Since $m_{\rm J} \propto T^{3/2} \rho^{-1/2}$ and $T$ is constant in an isothermal gas, $p(m_{\rm J})$ is directly related to the density probability distribution function (PDF). Previous simulations have shown that in isothermal, supersonically turbulent gas this is log-normal 
\citep[see e.g.][]{padoan97,klessen00,ostriker01,li2003,fed08,fed10}, and that the variance of the logarithmic density contrast $s \equiv \ln(\rho / \rho_{0})$ scales as \citep{molina12}
\begin{equation}
\sigma_{s}^2 = \ln \left[1 + b^{2} {\cal M}^{2} \frac{\beta}{\beta + 1} \right],
\end{equation}
where $1/3 \leq b \leq 1$ is a parameter related to the mix of solenoidal and compressive modes in the turbulent velocity field \citep{fed08,fed10}, ${\cal M}$ is the rms Mach number of the turbulence and $\beta = p_{\rm th} / p_{\rm mag}$ is the mean ratio of the thermal and magnetic pressures in the gas.  
Increasing the velocity dispersion of the gas broadens the density distribution and hence also the probability distribution for $m_{\rm J}$. It therefore leads to an increase in the   number of low-mass objects forming in the cloud (see e.g.\ Figure 1 in \citealt{padoan2002}). 

The \citet{pressschechter1974} formalism, originally developed to describe the mass function of collapsing dark matter halos, and first used to describe star formation processes in \citet{inutsuka2001}, is the starting point for \citet{hennebelle2008}, who derive an analytical expression for the distribution of the collapsing dense cores in a turbulent gas, leading to the CMF/IMF. In contrast with \citet{padoan2002}, reaching a pre-determined density threshold is not enough to ensure gravitational collapse in the \citet{hennebelle2008} model, since thermal and turbulent support on the scale of the dense cores play an important role. Taking all these aspects into account, \citet{hennebelle2008} derive the following equation for the distribution of collapsing cores\footnote{See the original paper for details of the derivation of this expression.}:
\begin{eqnarray}
\mathcal{N}(\tilde{M}) & = & 2\mathcal{N}_0 \frac{1}{\tilde{R}^{6}} \frac{1+(1-\eta)\mathcal{M}^2_\ast\tilde{R}^{2\eta}}{1+(2\eta+1)\mathcal{M}^2_\ast\tilde{R}^{2\eta}} \\
& & \times\left(\frac{\tilde{M}}{\tilde{R}^3}\right)^{A} \times\frac{\exp{(-\sigma_{s}^2/8)}}{\sqrt{2\pi}\sigma_{s}},
\label{eqhc2008}
\end{eqnarray}
where
\begin{equation}
A = -\frac{3}{2} - \frac{\sigma_{s}^2}{2} \ln{\left(\frac{\tilde{M}}{\tilde{R}^3} \right)},
\end{equation}
and where $\tilde{M}$ is the mass of the core in units of the mean Jeans mass of the cloud, $\tilde{R}$ is the size of the cloud in units of the mean Jeans length, $\eta$ is the power-law slope of the linewidth-size relationship, $\mathcal{N}_{0} \equiv \bar{\rho} / M_{\rm J}^{0}$ where $\bar{\rho}$ is the mean density of the cloud and $M_{\rm J}^{0}$ is the mean Jeans mass. 

Equation~\ref{eqhc2008} depends on the Mach number of the turbulence in two ways: explicitly via $\mathcal{M}_\ast$, the Mach number at the scale of a Jeans length, and implicitly via $\sigma_{\rm s}$, which as we have already seen depends on the Mach number on the scale of the entire cloud $\mathcal{M}$. While $\mathcal{M}_\ast$ increases with decreasing density, $\mathcal{M}$ increases with increasing size of the cloud due to Larson's scaling relations \citep{larson81}. For the sake of our study, we are interested in the behaviour of the protostellar mass distribution for the case in which both the density and the size of the cloud are constant but the velocity dispersion increases. This corresponds to varying both $\mathcal{M}_\ast$ and $\mathcal{M}$ but keeping their ratio fixed. In this case, increasing $\mathcal{M}$ leads to a shift of the peak of the IMF towards lower masses (see Figure 3 in \citealt{hennebelle2008}).

\citet{hopkins2012} noted that the results obtained in \citet{hennebelle2008} do not take into account that a gas cloud can be bound on several different scales, i.e. gas which is bound on a scale $R$ can still fragment into several other bound objects, and so on and so forth. The so-called first crossing distribution is the mass spectrum at the largest scale at which the gas is bound, representing giant molecular clouds. The IMF would then be described by the last crossing distribution. For the position of the CMF/IMF peak, \citet{hopkins2012} derives the following equation:
\begin{equation}
 M_{\rm sonic}\approx\left(\frac{4\sqrt{2}\pi}{3}\right)\mathcal{M}_h^{-2p/(p-1)}M_0,
 \label{eqh2012}
\end{equation}
where $p$ is the exponent of the turbulent velocity spectrum, $\mathcal{M}_{h}$ is the Mach number measured on large scales, and $M_{0}$ is the mass of the star-forming cloud. Equation~\ref{eqh2012} again confirms the trend of a decreasing turn-over mass with increasing Mach number (see Figure 3 in \citealt{hopkins2012}). 

Despite their different approaches, all of these theories study star formation in an isothermal gas under the influence of turbulence, self-gravity and a magnetic field and they all come to the conclusion that increasing the velocity dispersion of the gas, while keeping the size and the initial density of the cloud fixed, leads to more fragmentation and thus to the formation of smaller dense cores, producing a shift of the CMF peak towards lower masses. If the CMF and the IMF are self-similar, the same result also holds for the mass distribution of stars, although this assumption has been strongly debated in the literature. Indeed, it has been claimed that dense cores might fragment into numerous single stars, rather than one, or that they might not collapse at all (see e.g.\ \citealt{goodwin2004a}, \citealt{goodwin2004b}, \citealt{clark2005}, \citealt{goodwin2006}, \citealt{holman2013}, \citealt{klessenglover2014} and references therein). In particular, \citet{clark2007} suggests that due to the linear correlation between free-fall time and Jeans mass, in a given time many more low-mass than high-mass stars should form (the so-called `timescale problem'), thus leading to a more bottom-heavy shape for the IMF in comparison to the CMF.

More recently, \citet{hopkins2013}, \citet{guszejnovhopkins2015}, and \citet{guszejnovhopkins2016} extended the theory of \citet{hopkins2012} first allowing EOS different than isothermal and later developing a semi-analytical treatment that allows them to follow the fragmentation of the gas from the scale of giant molecular clouds to that of protostellar objects \citep[see][]{guszejnovhopkins2016}. Since in our simulations the EOS of the gas is isothermal in the regime in which we can resolve the self-gravitating gas, we will refer, when discussing our results, only to the analytical theories introduced in the beginning of this section. In order to compare to the results of \citet{guszejnovhopkins2016}, further numerical experiments would be needed that study the effect of varying EOS on the shape of the IMF.

Other interesting results were obtained by \citet{federrath2012} and \citet{federrath2013}, who summarise and extend the work by \citet{krumholz2005}, \citet{padoan2011}, and \citet{hennebelle2011} about the star formation rate (SFR) in a magneto-turbulent environment. They come to the conclusion that, given a system in virial equilibrium, the parameter having the most significant influence on the SFR is the supersonic Mach number, with an increase of the Mach number yielding an enhancement in the SFR of the cloud.

\subsection{This study}
The purpose of the present work is to run smoothed particle hydrodynamics (SPH) simulations characterised by different levels of turbulence, in order to investigate the influence of variations of the Mach number on the IMF and to test the predictions of the analytical theories presented above. In Section~\ref{sec:methods} we explain the methods used to conduct our study. Section~\ref{sec:sim} presents the results of our simulations, which are then discussed in Section~\ref{sec:dis}. In Section~\ref{sec:cave}, we list a few important caveats. Finally, in Section~\ref{sec:concl} we list the conclusions that we derive from our investigation. 

\section{Methods}
\label{sec:methods}
In this work we present the results of a set of numerical simulations that we perform using the smoothed particle hydrodynamics (SPH) code {\sc Gadget 2} \citep{springel2005}. In the first set of simulations, sampling the low-density regime, we simulate part of a dense molecular cloud in the form of a periodic box of $10$ pc in length, with a particle number density of $100\,\rmn{cm}^{-3}$ and solar metallicity (mean molecular weight $\mu=2.33$), corresponding to a total mass of $5750\,\rmn{M}_{\odot}$. We choose these parameters to match the conditions in nearby star forming regions (see e.g. \citealt{bergintafalla2007}). Using the equation from \citet{bateburkert1997}:
 \begin{equation}
  m_{\rm res}=\frac{M_{\rm box}}{N}\times N_{\rm neigh},
 \end{equation}
where $M_{\rm box}$ is the total mass of the box and $N_{\rm neigh}=57$ the number of neighbours used to compute the smoothing length of each SPH particle, we are able to calculate the mass resolution of the simulations. The total number of SPH particles contained in the box is $N=271^3=19\,902\,511$, and so $m_{\rm res}\approx 0.03 \, \rmn{M}_{\odot}$. In the second set of simulations, sampling the high-density regime, we keep the number of particles constant, but we shrink the box to a size of 3 pc and a mass of $\sim516 \, \rmn{M}_{\odot}$, corresponding to a mass resolution of $m_{\rm res}\approx 0.003 \, \rmn{M}_{\odot}$ (the key parameters for the different runs are summarised in Table 1). 
\begin{table*}
\centering
 \caption{Key parameters for the different runs.}
\begin{tabular}{@{}cccccccc@{}}
  Box size [pc] & Mass [${\rm M}_{\odot}$] & Particle number & Mass resolution [${\rm M}_{\odot}$] & $\mathcal{M}$ & SFE [\%] & Number of sinks & Simulation time [Myr] \\
\hline
  10 & 5750 & 19902511 & 0.029 & 1 & 8.84 & 327 & 89.5\\
  10 & 5750 & 19902511 & 0.029 & 5 & 9.03 & 378 & 12.5\\
  10 & 5750 & 19902511 & 0.029 & 10 & 8.93 & 366 & 6.4\\
  10 & 5750 & 19902511 & 0.029 & 15 & 8.81 & 278 & 7.9\\
  10 & 5750 & 19902511 & 0.029 & 20 & 8.82 & 299 & 7.2\\
  3 & 516 & 19902511 & 0.0026 & 5.5 & 4.89 & 120 & 4.68\\
  3 & 516 & 19902511 & 0.0026 & 8 & 4.88 & 84 & 4.53\\
  3 & 516 & 19902511 & 0.0026 & 11 & 4.85 & 123 & 4.04\\
\hline
\end{tabular}
\end{table*}

\subsection{The equation of state}
\label{sec:eos}
The mass resolution $m_{\rm res}$ presented in the previous paragraph is one of the key parameters in our simulations. As argued by \citet{bateburkert1997}, the local Jeans mass in SPH simulations should never be allowed to fall below the mass resolution, in order to ensure the correct fragmentation behaviour of the gas. We therefore adopt an EOS similar to the one presented in \citet{bateetal2003}. It is isothermal at low densities, with a temperature $T = 10$~K, but becomes polytropic at densities higher than the resolution density $\rho_{\rm res}$:
 \begin{equation}
  T\propto\rho^{\gamma-1}\,\rmn{with}\,
  \left\{
  \begin{array}{lr}
   \gamma=1  & \hspace{.5in} \rho\le\rho_{\rm res}\\
   \gamma=1.4 & \rho>\rho_{\rm res}
  \end{array}
\right.
 \end{equation}
The resolution density is calculated assuming $m_{\rm res}$ to be the smallest Jeans mass that can be resolved:
 \begin{equation}
  \rho_{\rm res}=\frac{3}{4}\frac{1}{\pi m_{\rm res}^2}\left(\frac{5}{2}\frac{k_BT}{G\mu m_p}\right)^3,
  \label{rho_res}
 \end{equation}
 where $k_{B}$ is Boltzmann's constant and $m_{p}$ is the proton mass.
 For the low-density regime, this corresponds to $1.6 \times 10^{-16}\rmn{g \, cm^{-3}}$, while for the simulations in the high-density regime we find $1.6 \times 10^{-14}\rmn{g \,cm^{-3}}$. Combining Equation~\ref{rho_res} with the EOS shows that a polytropic index $\gamma=1.4$ corresponds to an approximately constant Jeans mass, thus preventing the gas from fragmenting down to masses that cannot be resolved. Of course, in nature such a limit does not exist and fragmentation stops only when the opacity limit is reached, at $\rho_{\rm thick} \sim10^{-13}\rmn{g \, cm^{-3}}$ (see e.g.\  \citealt{low1976}, \citealt{bateetal2003}, \citealt{masunaga2000}). Our fragmentation limits lie at densities that are respectively $\sim3$ and $\sim1$ orders of magnitude lower than the opacity limit. We will discuss the effects of this choice on our results in Section~\ref{sec:hr}.  

\subsection[]{Sink particles}
In {\sc Gadget 2}, the size of the computational time steps depends on the smoothing length of the SPH particles which, in turn, is a function of their density. Thus, when the gas collapses and the density increases, the time steps become shorter and the code becomes increasingly slower. It is therefore impossible to follow the collapse of gas to protostellar densities while also following the large-scale evolution of the cloud for an extended period. To avoid this problem and to allow us to follow the evolution of the cloud over multiple dynamical times, we insert `sink' particles (as originally introduced by \citealt{bate1995}) in place of the SPH particles whose density exceeds a fixed critical value, using the implementation of \citet{jappsen2005}. We choose the critical density to be two orders of magnitude higher than the resolution density, which yields $\rho_{\rm crit}=1.6 \times 10^{-14}\rmn{g \, cm^{-3}}$ for the low-density regime and $\rho_{\rm crit}=1.6 \times 10^{-12}\rmn{g \, cm^{-3}}$ for the high-density regime.

A sink particle is created from a given SPH particle with $\rho>\rho_{\rm crit}$ and smoothing length $h_{\rm crit}$ smaller than the Jeans length at $\rho_{\rm crit}$. The sink immediately accretes all the SPH particles lying within a sphere of radius $h_{\rm crit}$, inheriting their total mass and momentum. The formation of a sink particle, though, is not immediate. The candidate particles must fulfil several criteria: the central particle must have a smoothing length $h<h_{\rm crit}$, while the companions, in order to be accreted, must lie on the same integrational time step, be bound, have negative $\nabla\cdot\vec{v}$ and $\nabla\cdot\vec{a}$ and the closest sink particle must be at a distance larger than $h_{\rm crit}$. 

After their creation, the sinks can interact with SPH particles and other sinks only gravitationally. The accretion of further SPH particles is regulated by two further parameters, the inner and outer accretion radii, $r_{\rm in}$ and $r_{\rm out}$. If a SPH particle is closer to a sink than $r_{\rm out}$, is gravitationally bound to it, and is more bound to it than to any other sink, then it gets accreted by that sink. When an SPH particle lies within $r_{\rm in}\approx100\,\rmn{AU}$ of the nearest sink, it is automatically accreted by that sink without any further tests.

If we were to follow the evolution of the gas down to the opacity limit, the gas from which sink particles form in the low-density regime would be likely to form an accretion disc around a central object that would then fragment further into multiple stellar systems. Thus, when talking about sinks in the low-density regime we refer to dense regions that would probably collapse into small stellar clusters. In the simulations with a higher mass resolution, we can follow the collapse of the gas to much higher densities, and we observe the formation and fragmentation of accretion discs as described above. Consequently the units represented by the sink particles in the high-density regime can reasonably be interpreted as protostellar objects, as the resolution density and hence the limit for fragmentation is only 1 order of magnitude lower than the opacity limit. In both cases we refer to the sink particles mass function as system mass function. We will discuss this issue in more detail in  Section~\ref{sec:hr}.

\subsection{Driven turbulence}
We aim to study the influence of supersonic turbulence on the star formation process. In nature, turbulence observed in molecular clouds can be driven in numerous ways, such as supernovae explosions, stellar winds, the shear of the galactic disc, etc. and naturally decays in the period between two consecutive driving events (see e.g. \citealt{maclowklessen2004}, \citealt{klessenglover2014}). For our numerical experiment, we decided to drive turbulence periodically on large scales (of the order of the box size). The SPH particles are distributed on a three-dimensional particle grid filling the box and are interpolated with a random velocity field at regular time intervals. 

The turbulence field is created in Fourier space as a random velocity field whose power spectrum is characterised by a top-hat function $P(k)\propto k^\alpha$ with $\alpha=0$ and $k=L/\lambda$ ranging from 1 to 2, where $L$ is the size of the box, $\lambda$ is the driving wavelength and $k$ a dimensionless wave number. Our choice therefore corresponds to large-scale driving.
Further details concerning the construction of the velocity field and the numerical implementation of this method in an SPH simulation can be found in \citet{maclowetal1998} and \citet{klessen2000}. In \citet{maclow1999}, it was shown that the turbulent kinetic energy in a supersonically turbulent cloud decays with time as
\begin{equation}
    \dot{E}_{\rm kin}=-\eta_v m\tilde{k}v_{\rm rms}^3,
\end{equation}\label{edot}
where $\eta_v=0.21/\pi$ is a constant, $m$ is the mass of the cloud, $v_{\rm rms}$ is the root mean square turbulent velocity and $\tilde{k}=(2\pi/L)k$ is the dimensionless driving wavenumber. Thus, after every time interval $\Delta t$ (which was chosen to be equal to a tenth of the free fall time, i.e. $\Delta t = t_{\rm ff}/10$) an amount of kinetic energy equal to $\dot{E}_{\rm kin} \times \Delta t$ is injected into the box. In each simulation, this process is repeated in the absence of self-gravity until the mean kinetic energy reaches a steady state. Afterwards, the gravitational force is switched on, but the kinetic energy is still injected at fixed time intervals in order to keep the Mach number approximately constant. 

\section{Results}
\label{sec:sim}
We run simulations characterised by several different Mach numbers $\mathcal{M}=v_{\rm rms}/c_s$, where $c_s$ is the sound speed, in order to determine the effect of variations in the turbulent state of a molecular cloud on star formation and especially on the mass distribution of the stars that form. The simulations in the low-density regime were stopped when a star formation efficiency (SFE) of $\sim 8-10 \%$ is reached, which is consistent with observations of star forming molecular clouds (see e.g.\ \citealt{maclowklessen2004}, \citealt{mckee2007}, \citealt{klessenglover2014}).

\subsection{Low-density regime}
\label{low_res}
\begin{figure*}
 \includegraphics[scale=0.14]{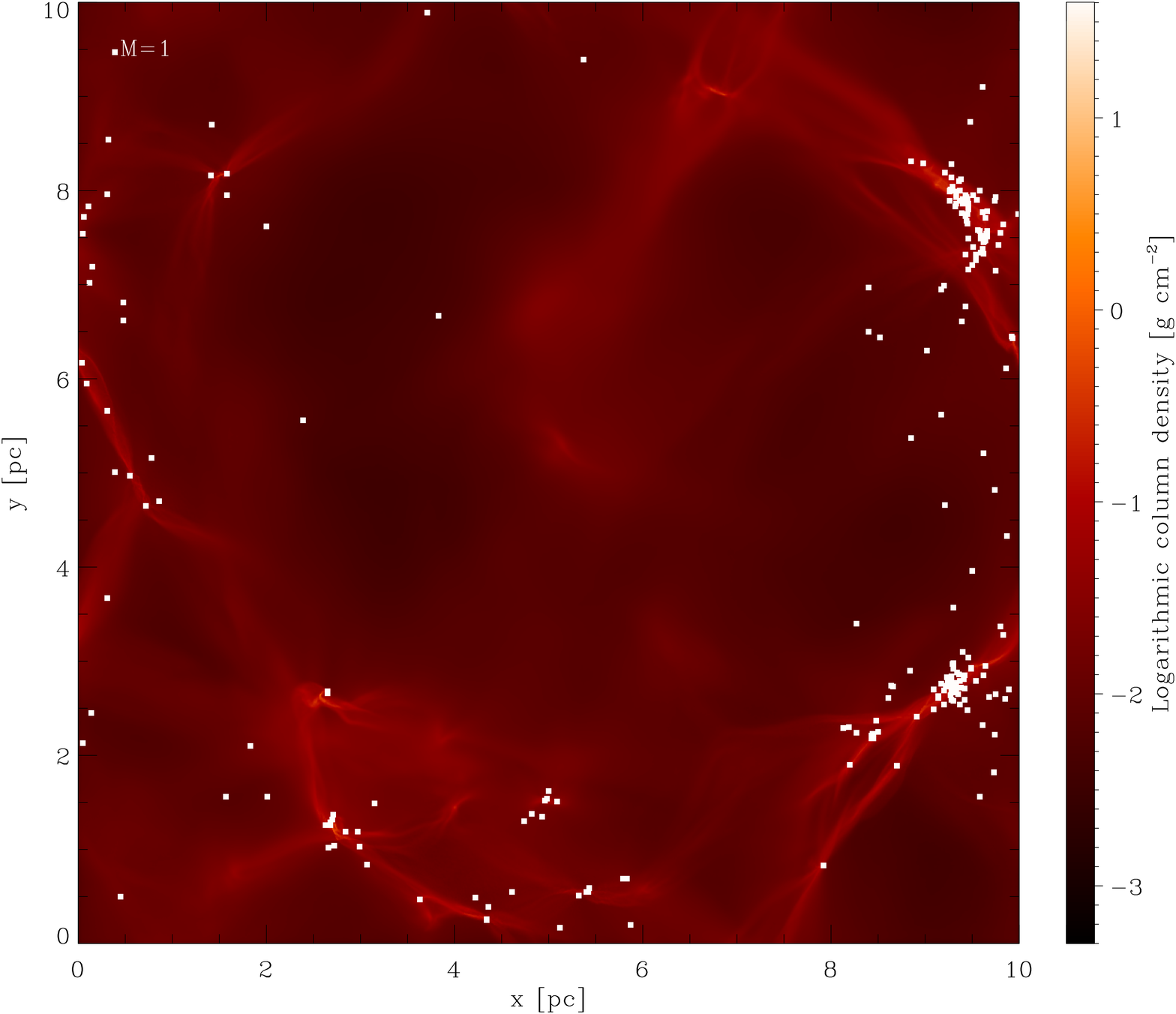}
 \includegraphics[scale=0.14]{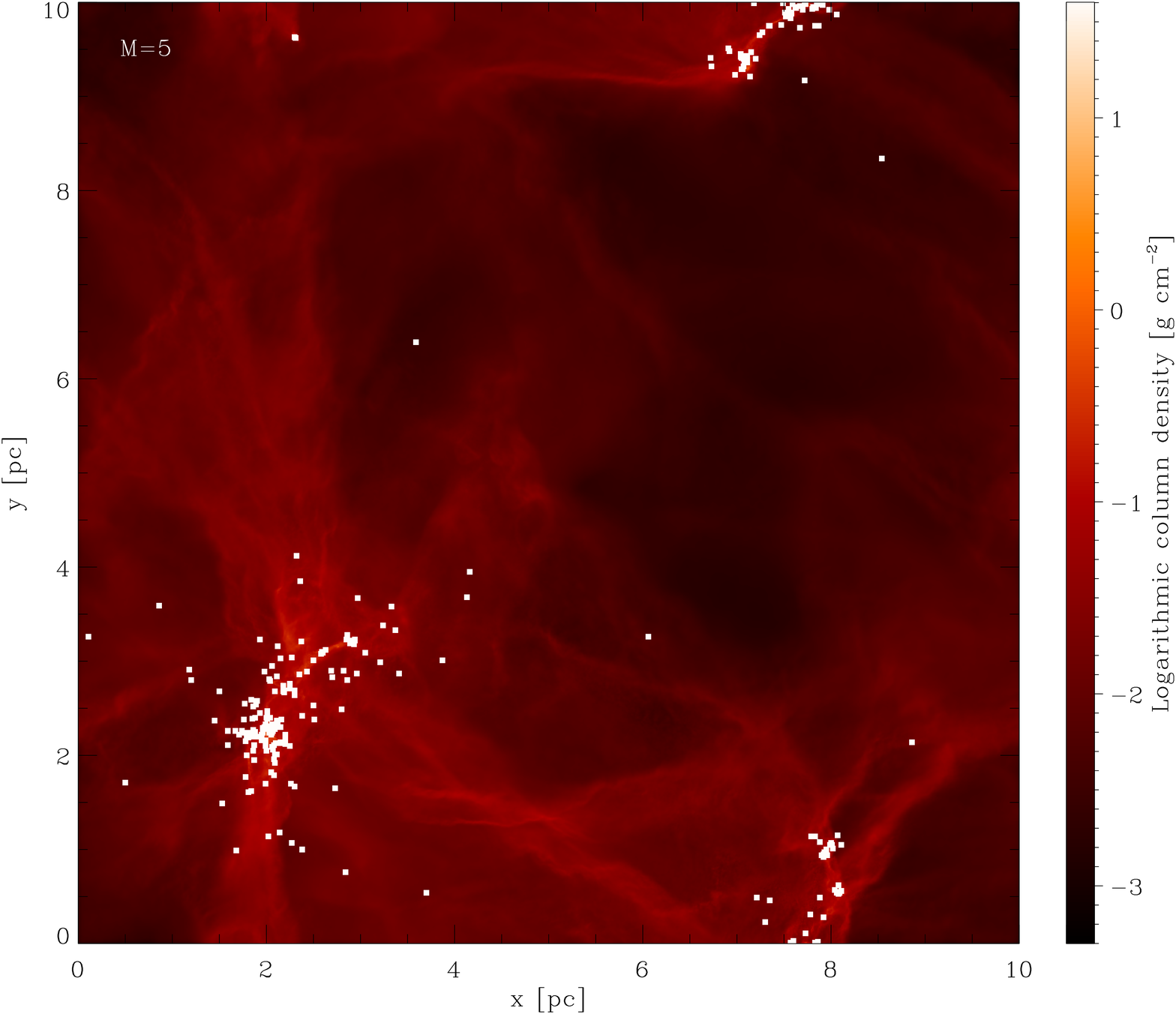}
 \includegraphics[scale=0.14]{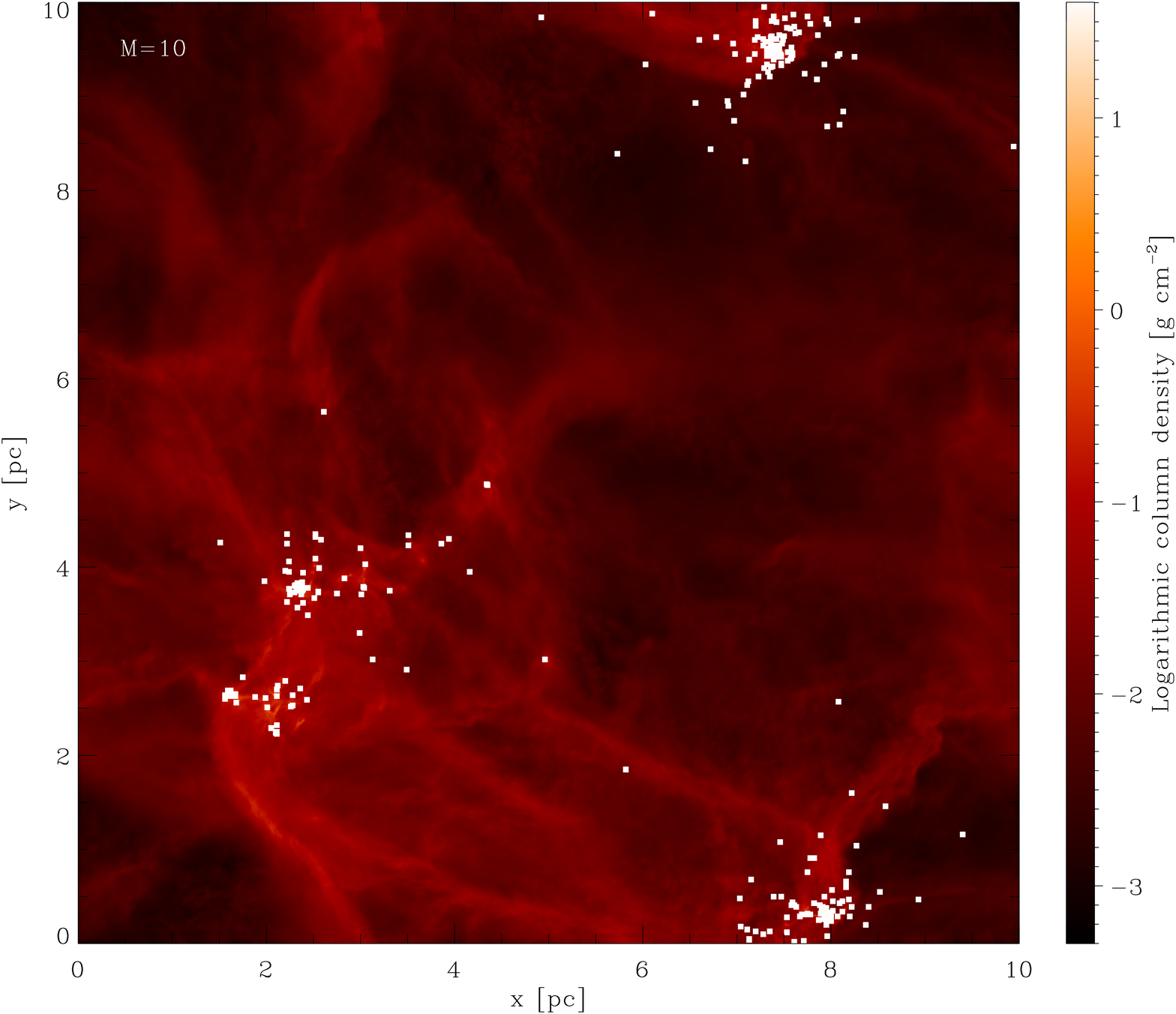}
 \includegraphics[scale=0.14]{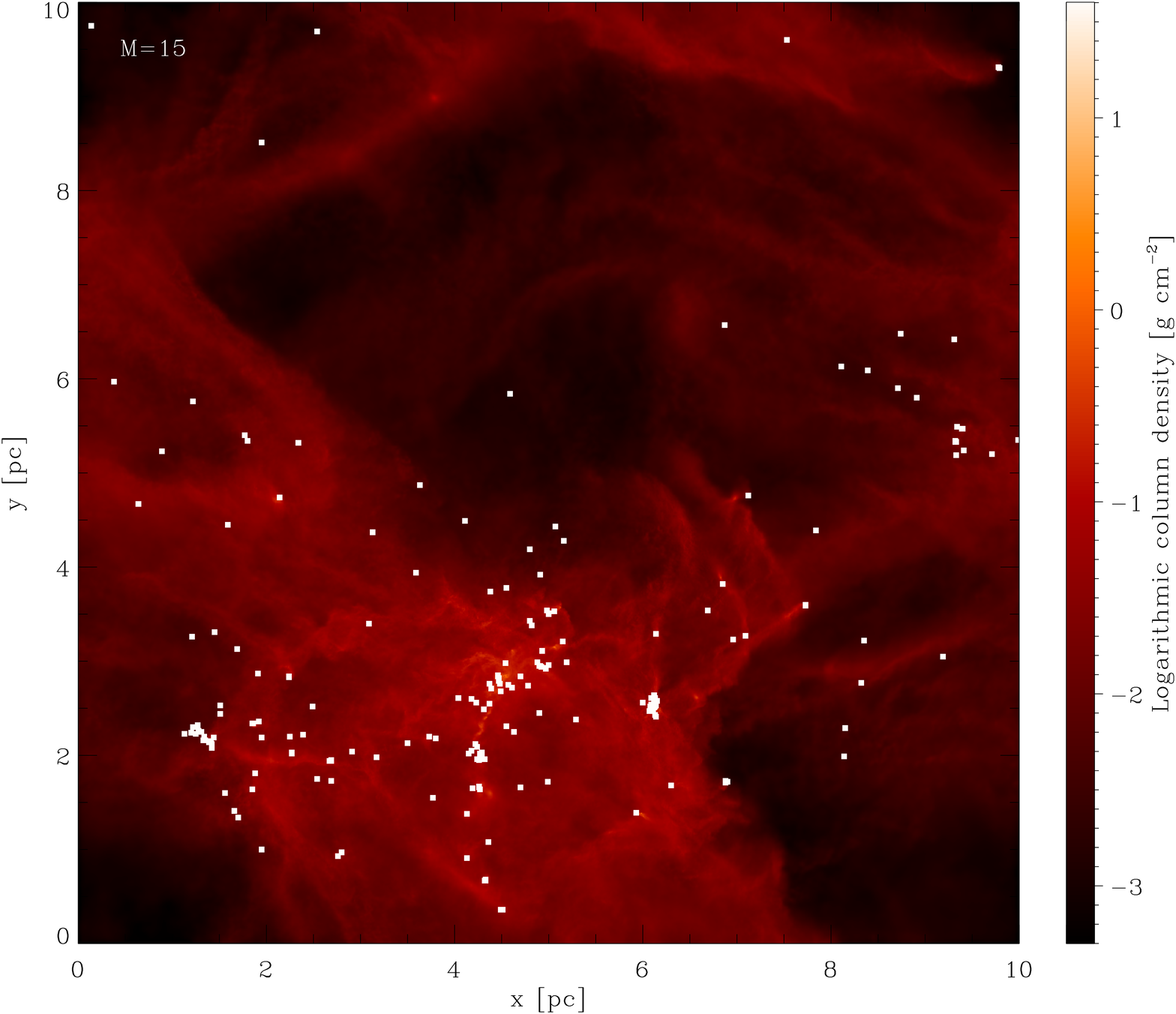}
 \includegraphics[scale=0.14]{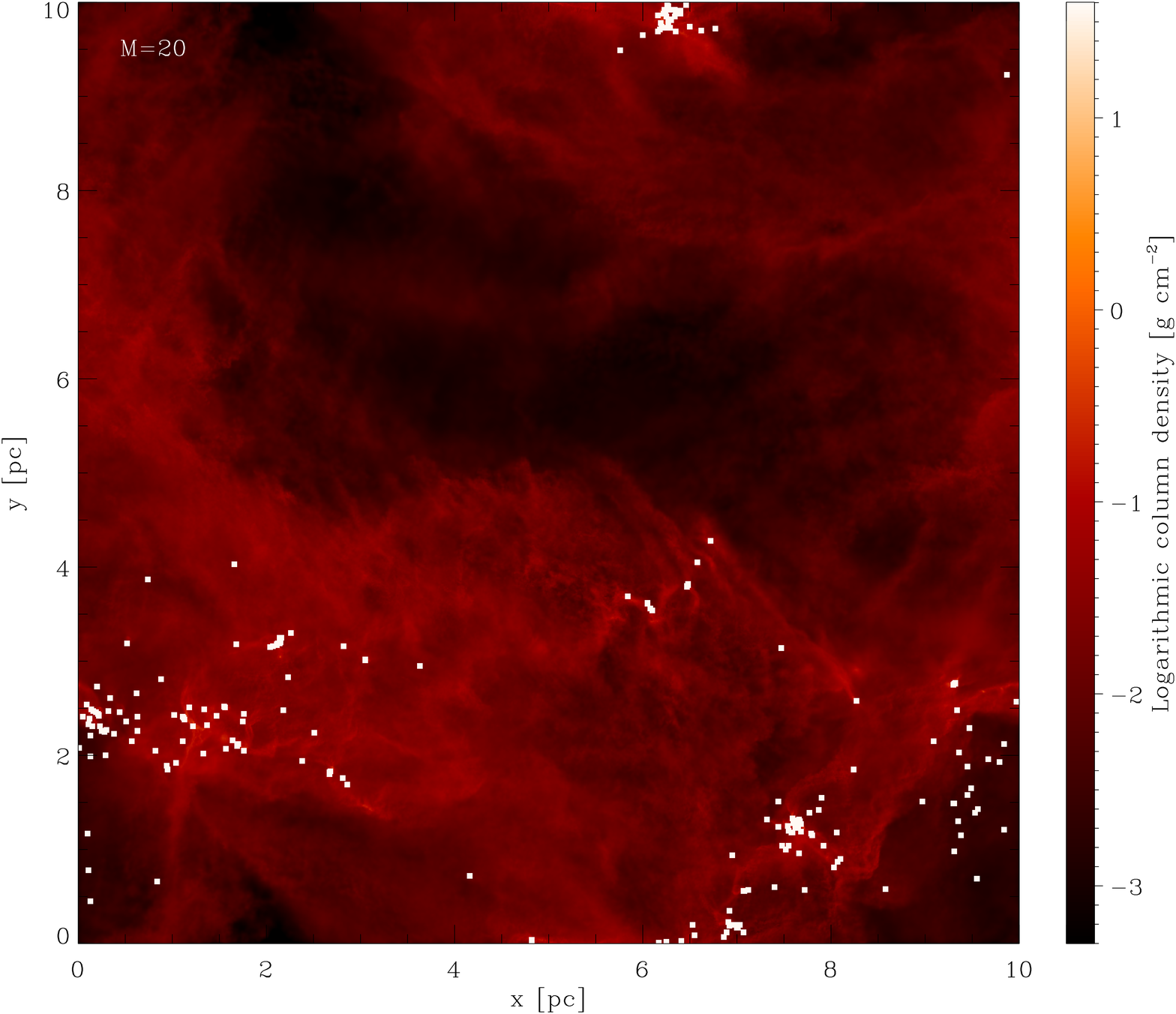}
\caption{Snapshots showing the distribution of gas and sink particles at the end of the simulations with lower mass resolution, when a total mass of $\sim500 \, {\rm M}_{\odot}$ is contained in protostellar objects. From the left to right and from top to bottom $\mathcal{M}=1,5,10,15,20$ and $t\approx89.5,12.5,6.4,7.9,7.2$ Myr. The filaments and structures of the gas that can be seen e.g. at $\mathcal{M}=1,5$ tend to be `washed out' when going to higher Mach numbers. Also the clustering of protostellar objects seems to be hindered by the increasingly high turbulence of the gas.}
 \label{fin_snap}
\end{figure*}

We begin by running a set of simulations with the initial conditions described in Section~\ref{sec:methods}. The Mach numbers $\mathcal{M}=1,5,10,15,20$ characterising the different runs are chosen within the ranges indicated by observational data (see e.g. \citealt{bergintafalla2007}). Figure~\ref{fin_snap} shows the final snapshot of each simulation. It is interesting to observe that the gas structure seems to become progressively more `washed out' as we increase the Mach number of the simulation. 
The well-defined filaments that can be seen in the simulations with $\mathcal{M}=1,5$ tend to get disrupted as we increase the velocity dispersion of the gas. Also the clustering behaviour of the sink particles produced in the different runs decreases as the velocity dispersion of the gas increases. 

\begin{figure*}
 \includegraphics[scale=0.3]{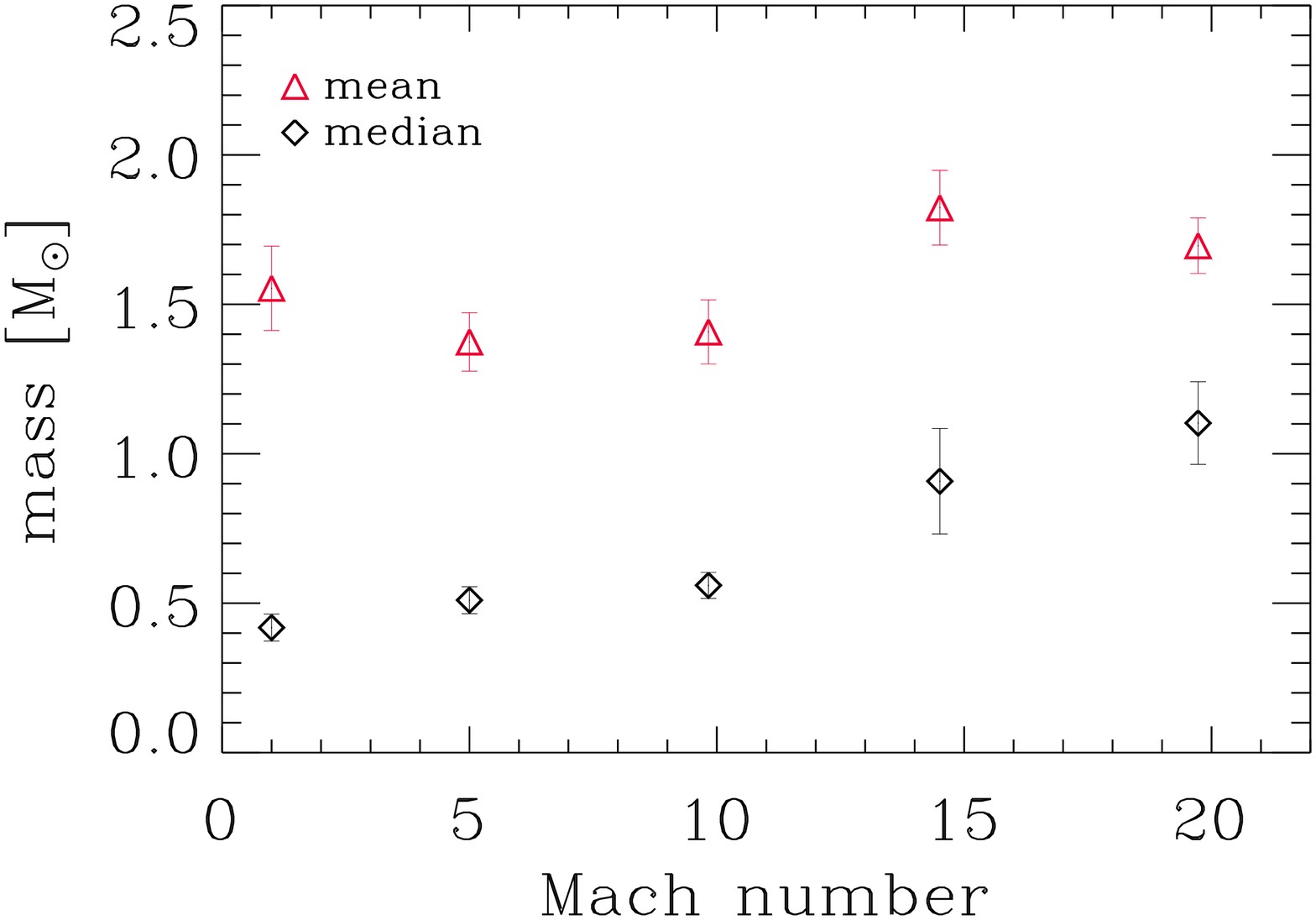}
 \includegraphics[scale=0.3]{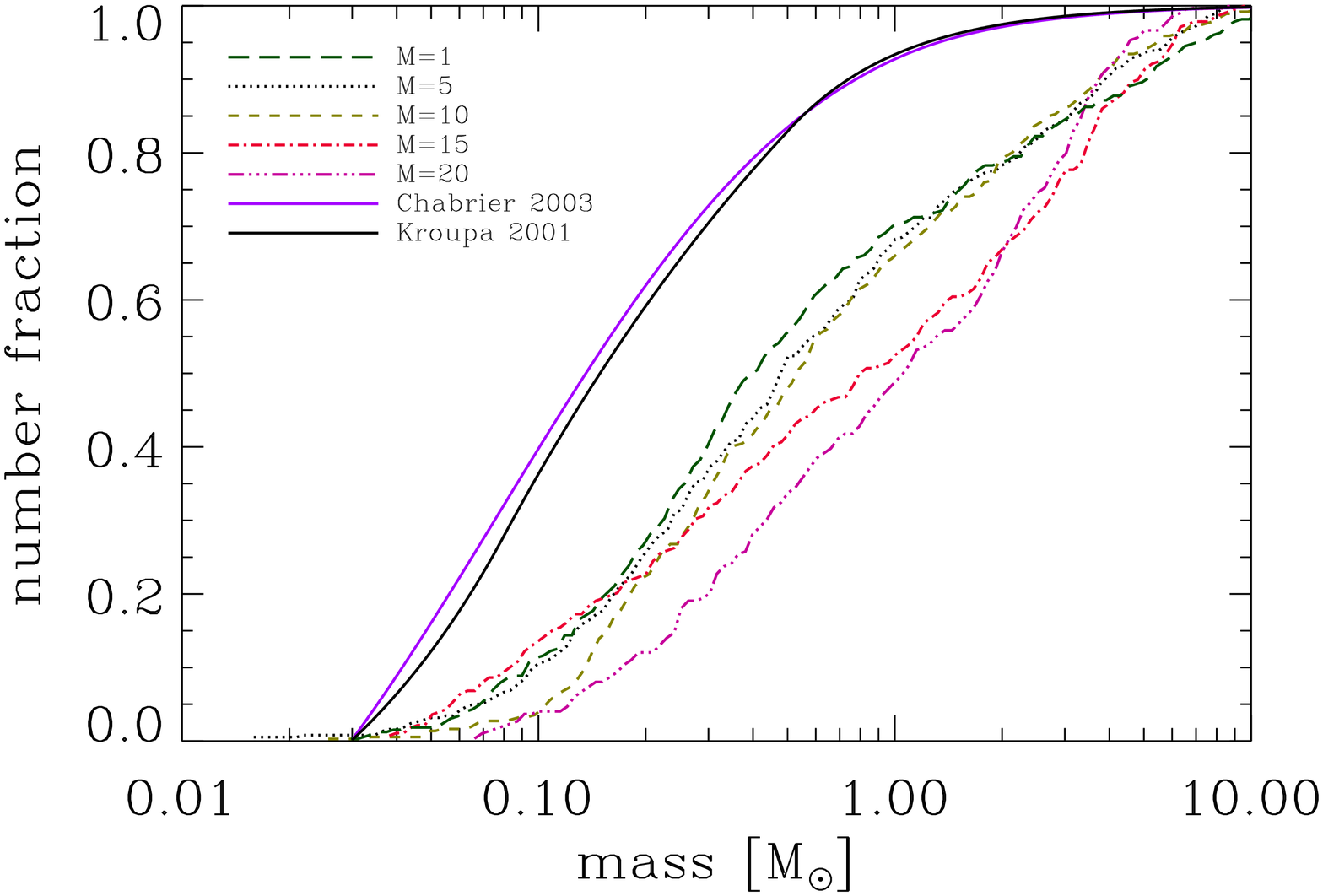}
\caption{\textit{Left panel:} Mean and median of the mass distributions obtained for the simulations in the low-density regime plotted as a function of the corresponding Mach number. \textit{Right panel:} Cumulative IMFs for the same simulations. The results are compared to a classical Chabrier and Kroupa IMF (\citealt{chabrier2003}, purple solid line, \citealt{kroupa2001}, black solid line)}.
 \label{res_driven}
\end{figure*}

The simulations presented here are very expensive in terms of computational resources and time. For each of the runs, 128 CPUs were employed for several weeks, and so we are restricted to only a handful of simulations. To characterise the statistical significance of our results, we perform a bootstrapping analysis. We resample the masses of the sink particles in each simulation 10000 times, and each time calculating mean and median mass of the distribution. In the end, we compute the mean and standard deviation for both quantities. The results are shown in Figure~\ref{res_driven}: in the left panel the mean and median (indicating the location of the peak of the IMF) of the mass distributions and their errors are plotted as a function of the corresponding Mach number, while in the right panel the distributions are shown in the form of cumulative IMFs. 
We observe that increasing the Mach number leads to a shift of the peak of the mass distribution at the end of the run towards higher masses. Taking into account the error ranges, this does not agree with the expectations of analytic theories which, we recall, predict an increasing amount of mass included into low-mass stars when increasing the velocity dispersion of the gas (see discussion in Section~\ref{sec:introth}).

Before we discuss the physics that may explain this trend, we first need to address another issue that is evident from the plots: none of the mass functions look like the `classical' IMF. Instead, all of them have a very small fraction of the total stellar mass contained in low-mass stars (see Figure~\ref{res_driven}, right panel), resulting in the IMFs being systematically too `flat' (i.e.\ too top-heavy). Regardless of the details of the relationship between the strength of the turbulence and the shape of the protostellar mass distribution, if it is not possible for our simulations to reproduce the IMF observed in the solar vicinity for any choice of turbulent initial conditions, we must be missing something important about the physics underlying the process of star formation. We will address this in the next section.

\subsection{High-density regime}
\label{sec:hr}
As we have already pointed out, the `flatness' of the mass distributions arising from the first set of simulations is systematic. Thus, before proceeding we must investigate the origin of this effect. 

In Section~\ref{sec:methods}, we discussed how we artificially prevent fragmentation from occurring at densities higher than the maximum resolvable density $\rho_{\rm res}$, in order to ensure that any fragmentation that we see is physical rather than numerical in nature. Similar behaviour occurs in nature once the gas becomes optically thick, but in our first set of simulations, $\rho_{\rm res}$ is $\sim3$ orders of magnitude smaller than the actual density at which the gas becomes optically thick, $\rho_{\rm thick}$. In these simulations, we therefore suppress any fragmentation occurring in gas with densities in the range $\rho_{\rm res} < \rho < \rho_{\rm thick}$. Since the Jeans mass in this regime is very small ($m_{\rm J} < m_{\rm res} \simeq 0.03 {\rm M_{\odot}}$), it is possible that this is responsible for the fact that our simulations underproduce low-mass stars, regardless of our choice of ${\cal M}$.

Ideally, this hypothesis could be tested by repeating the simulations with a higher mass resolution, thus catching the behaviour of the gas both on large (low density) and on small scales (high density). Maintaining the same initial conditions as described in Section~\ref{sec:methods}, but increasing the mass resolution e.g.\ by a factor 10 would imply an increase in the number of SPH particles by the same factor, i.e.\ from $\sim 2 \times 10^7$ to $\sim 2 \times 10^8$. Unfortunately, the computational effort required to carry out star formation simulations of this size for the desired number of dynamical times is 
too large to be practical given our current resources, meaning that another solution must be found. Therefore, rather than increasing the number of SPH particles, we keep it fixed and instead shrink the box to a size compatible with the mass resolution we want to reach. To this purpose, we use Larson's relations \citep{larson81} to `zoom' into a region within our clouds.

The new box has mass $516 \, \rmn{M}_\odot$ and size 3 pc, resulting in a mass resolution $m_{\rm res} = 0.0026 \, \rmn{M}_\odot$. 
The `knee' of the EOS (i.e.\ the point at which it steepens) is consequently shifted to a much higher density ($\rho_{\rm res}=1.6 \times 10^{-14} \rmn{g \, cm^{-3}}$, only one order of magnitude below the opacity limit). Consistency with Larson's relations also requires us to adjust the Mach numbers used in the simulation, so that the velocity dispersion on the scale of the box in each of our simulations in the high-density regime remains the same as the velocity dispersion on a scale of 3~pc in the corresponding simulation in the low-density regime. We therefore carry out simulations using 
$\mathcal{M}_{\rm new}=5.5,8,11$ corresponding to the previous $\mathcal{M}_{\rm old}=10,15,20$. Since the higher mass resolution of the simulations results also in a shift in the critical density for sink formation towards higher densities, the computational time steps become smaller. In addition, allowing fragmentation down to smaller masses considerably slows down the process of star formation and accretion. For these reasons we decided to run this set of simulations only until an SFE of $\sim5 \%$ is reached.

\begin{figure*}
  \includegraphics[scale=0.3]{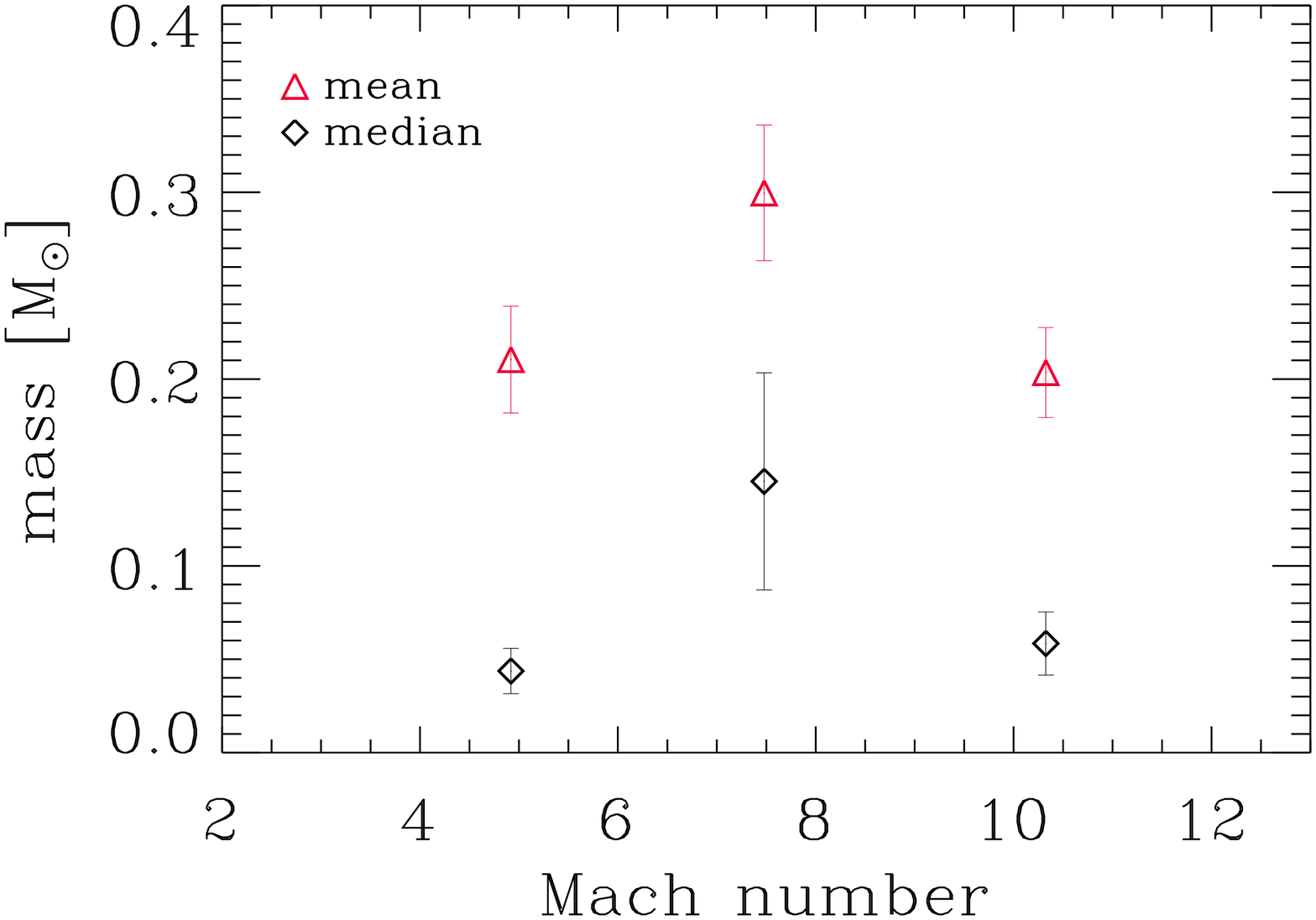}
 \includegraphics[scale=0.3]{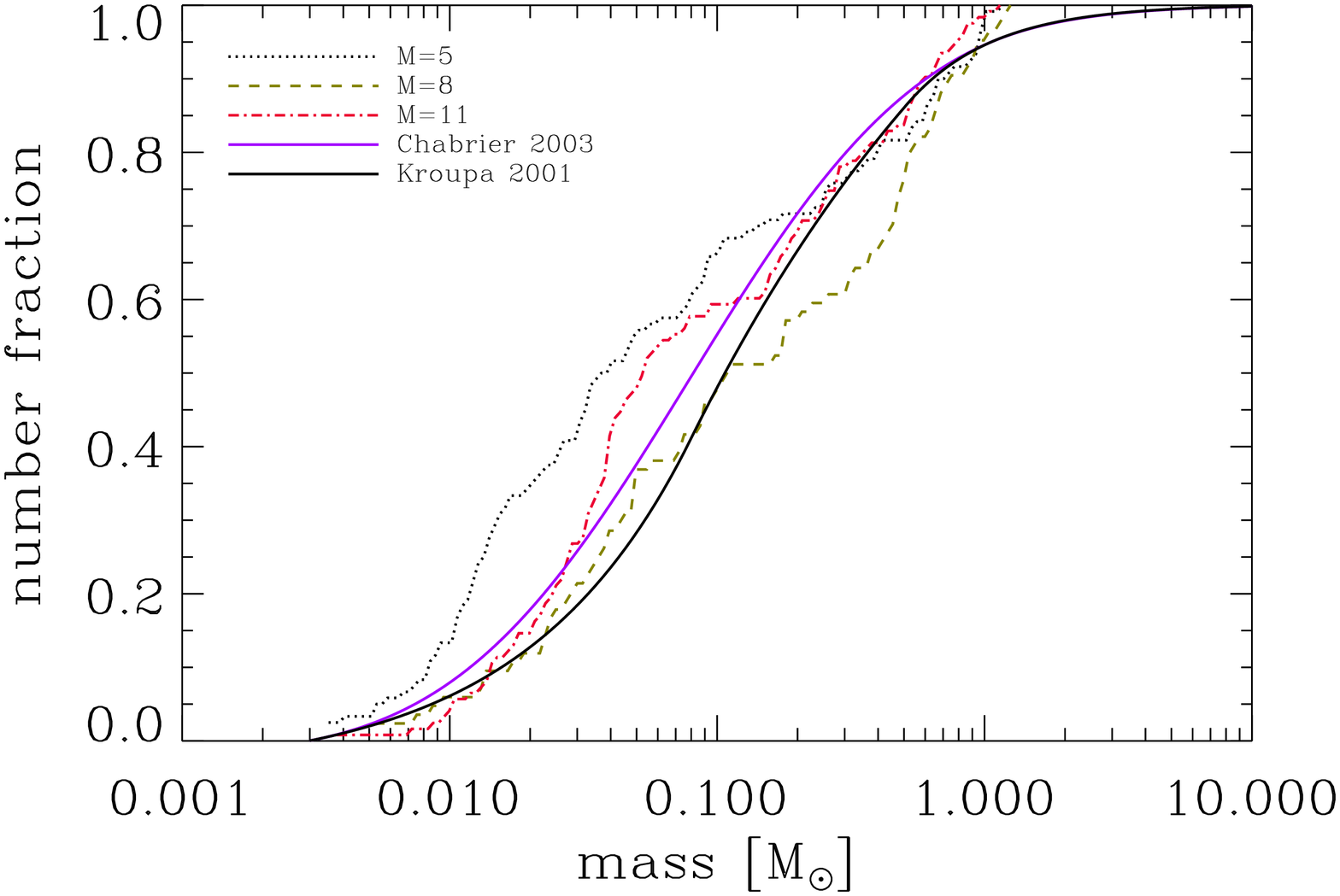}
\caption{\textit{Left panel:} Mean and median of the mass spectrum as a function of the corresponding Mach number for the simulations in the high-density regime. \textit{Right panel:} Cumulative IMFs resulting from the simulations with high mass resolution. The comparison with the Chabrier and Kroupa IMF shows that increasing the mass resolution of the simulations allows us to reproduce the IMF observed in the solar neighbourhood.}
\label{res_high}
\end{figure*}

The results of this new set of simulations are shown in Figure~\ref{res_high}, again analysed with the same method as described in Section~\ref{low_res}. This time, all the resulting IMFs lie within a small range around the curve represented by the Chabrier IMF. Indeed, the run with the lowest Mach number actually produces an even higher number of brown dwarfs than predicted by the Chabrier IMF. This confirms that the systematical lack of low mass objects obtained with the simulations at lower mass resolution was simply due to the artificially low critical density that we imposed for numerical reasons. Increasing the mass resolution allows us to increase the critical density, permitting the gas to fragment on smaller scales and opening up a new star-formation regime.

\begin{figure*}
 \includegraphics[scale=0.14]{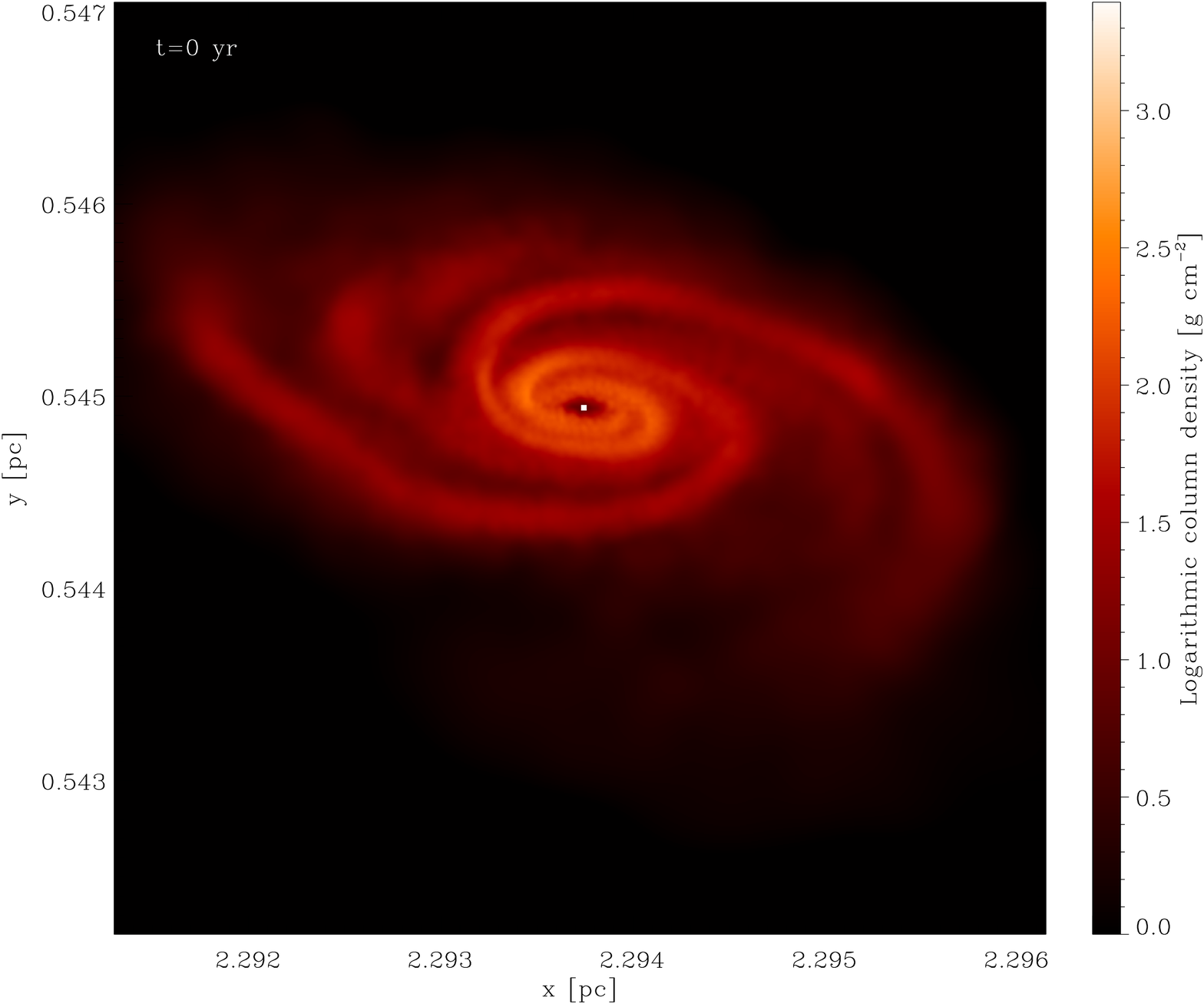}
 \includegraphics[scale=0.14]{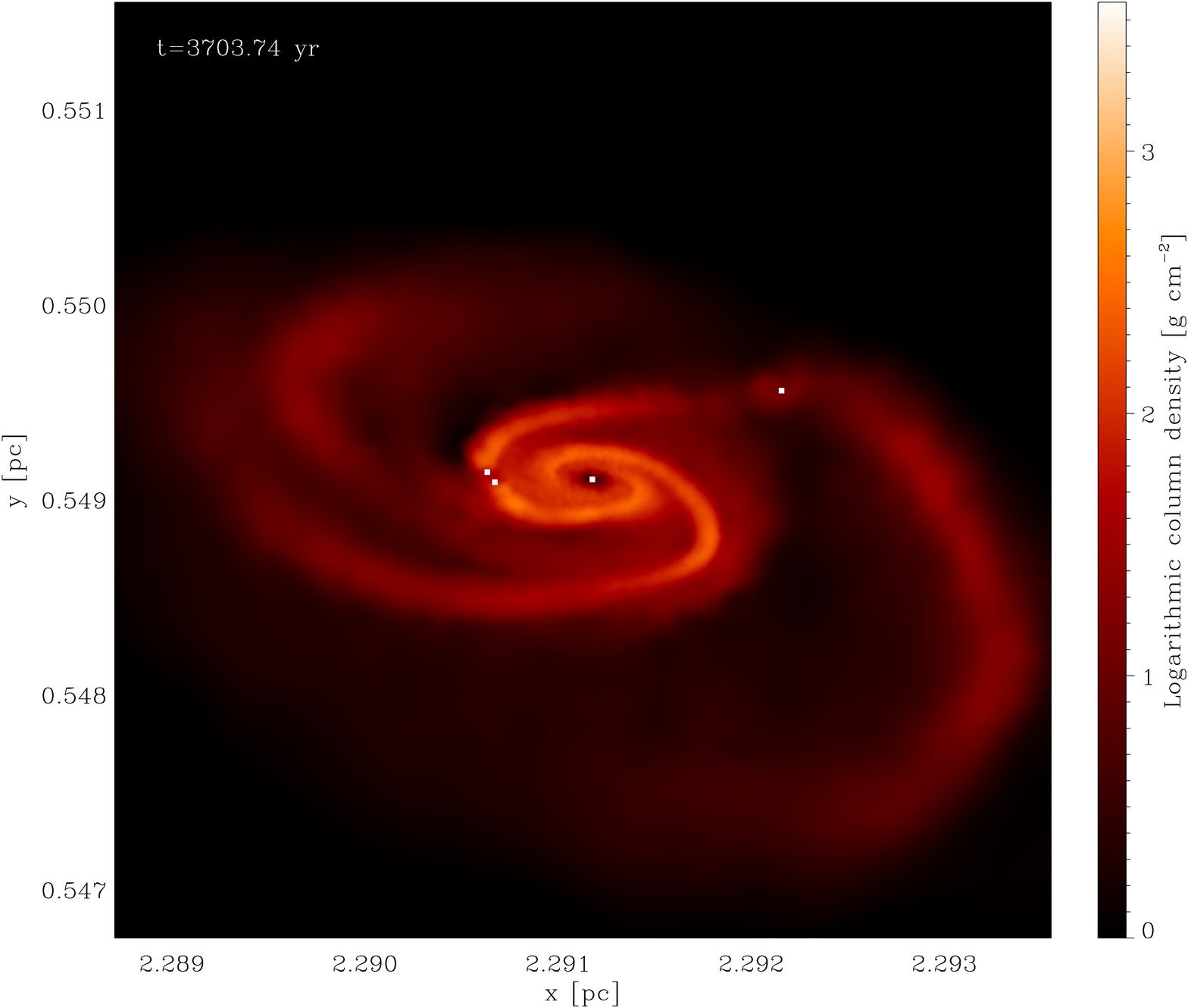}
 \includegraphics[scale=0.14]{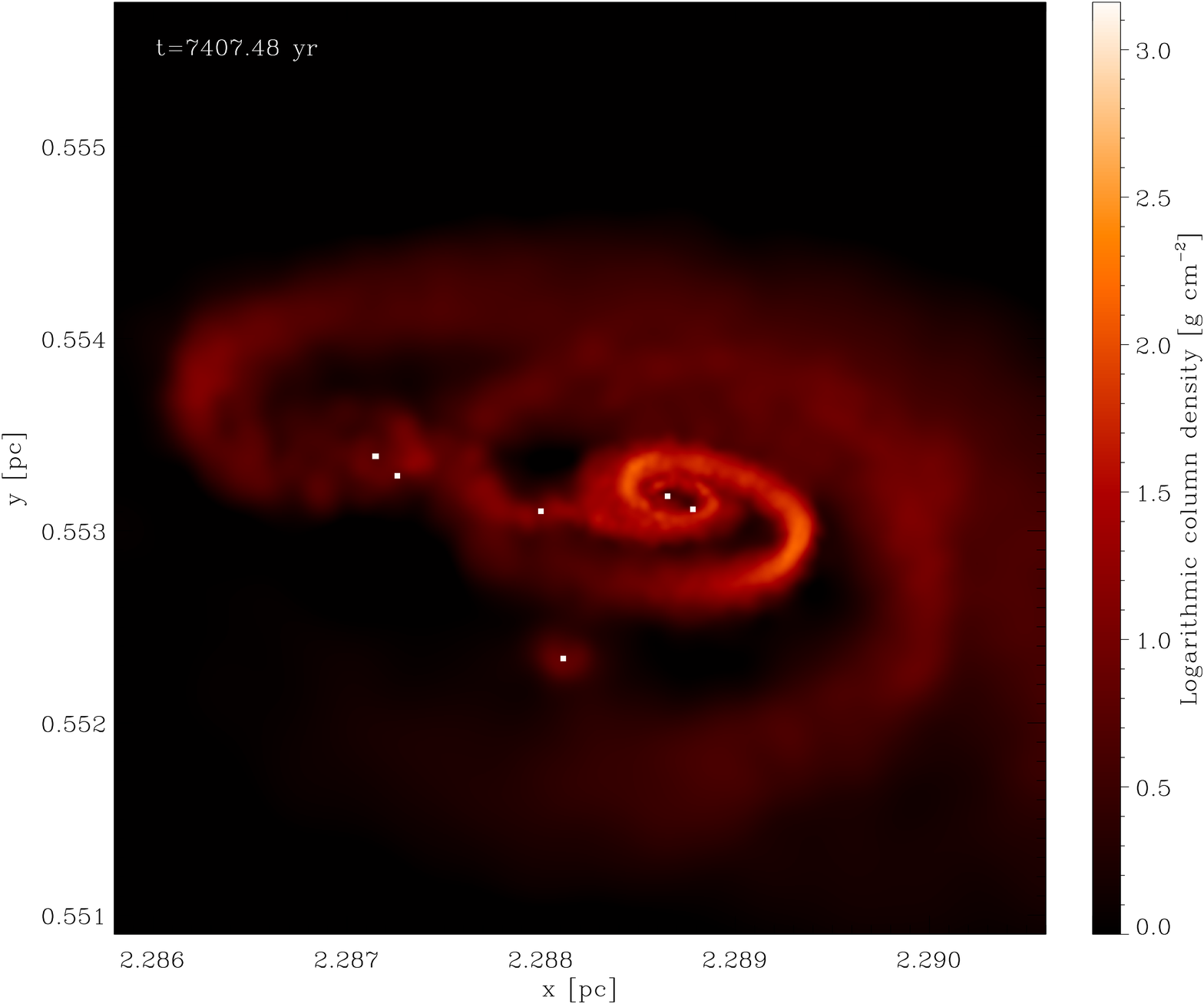}
 \includegraphics[scale=0.14]{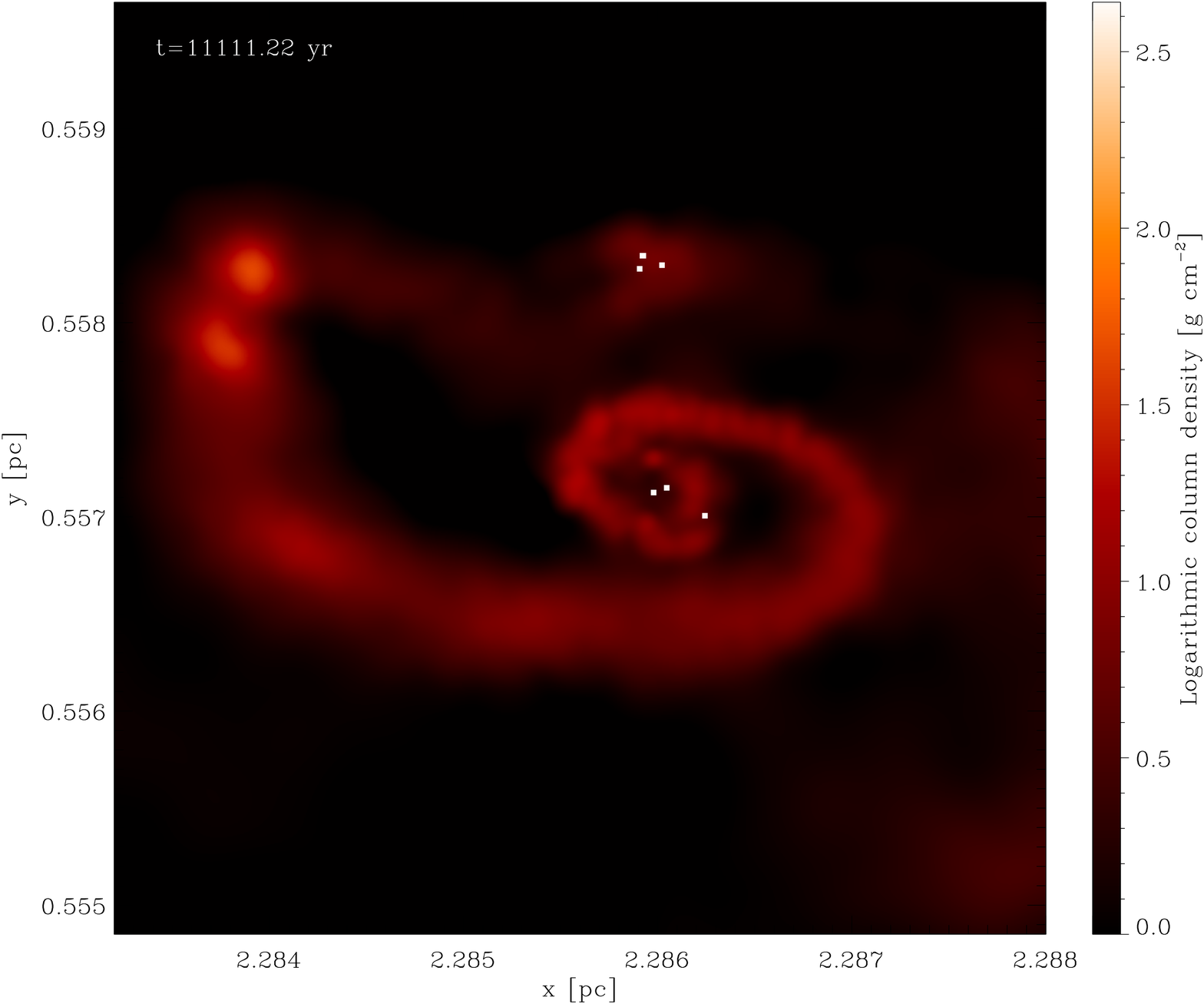}
 \includegraphics[scale=0.14]{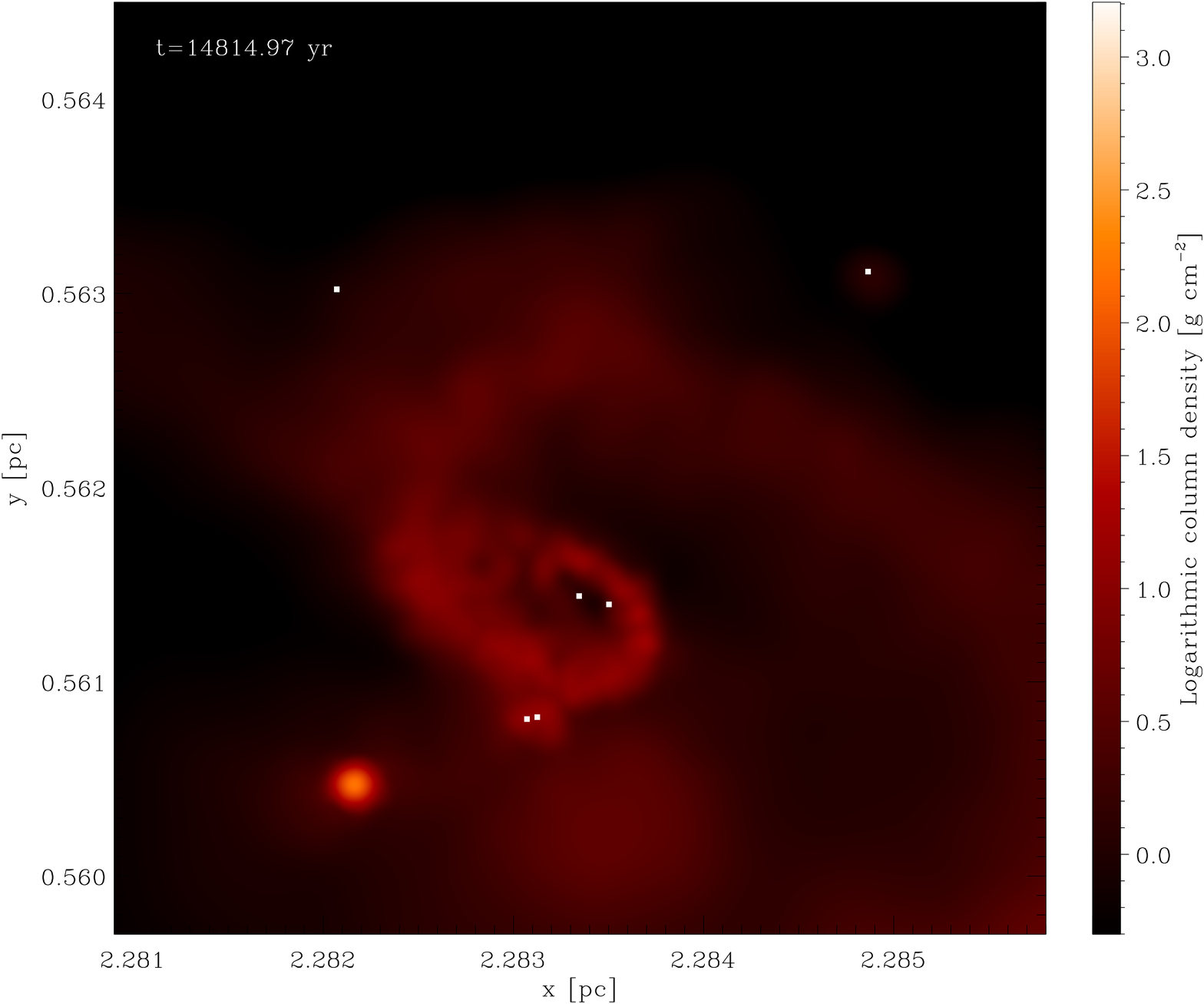}
 \includegraphics[scale=0.14]{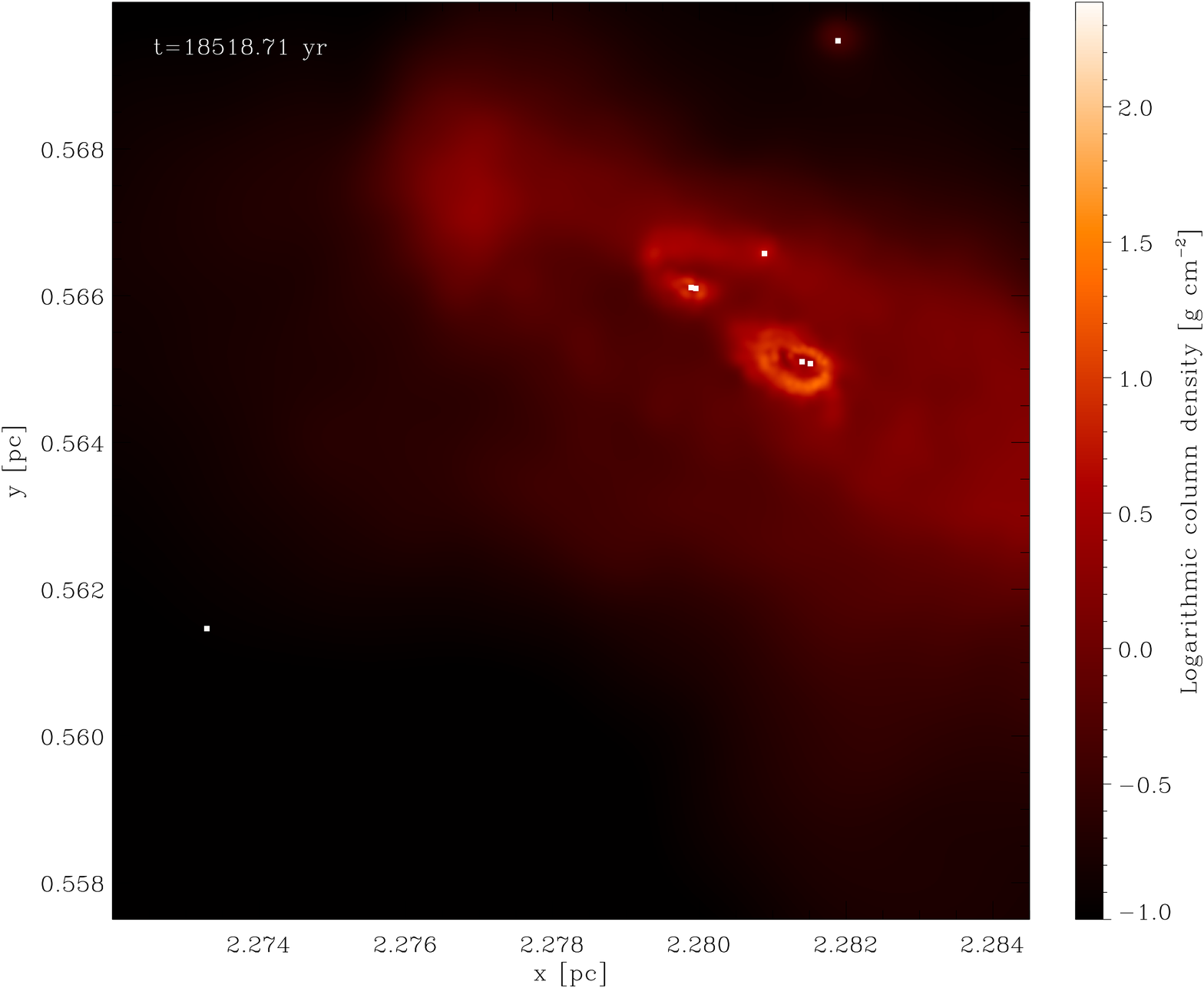}
  \caption{Consecutive snapshots showing the fragmentation and collapse of an accretion disc and consequent expulsion of low-mass objects from a multiple system in the high-density regime with $\mathcal{M}=8$ (ordered from left to right and top to bottom). In the runs with lower mass resolution the accretion radius of a sink particle is equal to $\sim 100$ AU ($\sim 10$ AU in the present run), which would cover most of the densest region in the top-left panel. In the low-density regime, such a dense core would have collapsed to form a single sink particle and accreted the surrounding gas.}
 \label{im_high}
\end{figure*}

Dense gas that would have collapsed into a single protostellar object in the low-density regime can now build an accretion disc that will fragment into many protostellar objects (see e.g.\ \citealt{bonnell1994}, \citealt{batebonnell1997}, \citealt{bateetal2003}, \citealt{meru2012}, \citealt{offner2014}).
Figure~\ref{im_high} illustrates this effect with a series of consecutive snapshots of a fragmenting accretion disc. The central region of this disc has roughly the size of the accretion radius used in the simulations with lower mass resolution, and would have consequently been substituted by a single sink particle. In the new set of simulations the disc can be resolved and fragmentation can take place.

The considerable time that elapses between two consecutive snapshots ($\sim3700$ yr) does not allow us to state with certainty if low-mass objects form exclusively through disc fragmentation, or if they also form in isolation. Studying the snapshots in which low-mass sinks appear for the first time we found some of them lying at large distances from the nearest high-mass object. However, the distances and velocities of the sinks are compatible with them having formed via disc fragmentation and subsequently been dynamically ejected from the system. This aspect should be better investigated by shortening the time between snapshots, in order to follow the formation of the low-mass objects and the evolution of their hosting systems with more accuracy, but this goes beyond the scope of this work.

We have shown that by increasing the mass resolution of the simulations we can recover the shape of the IMF observed in the solar neighbourhood. 
However, we also find that in this new set of simulations, the response of the IMF to changes in the velocity dispersion is different from the behaviour in the low-density regime. In the first set of simulations an increase in the velocity dispersion of the gas produces a shift of the mass distribution towards higher masses. On the other hand, in the new set of simulations, describing the high-density regime,  there appears to be no clear relationship between the Mach number of the gas and the resulting IMF. Importantly, in neither set of simulations do we see any evidence for the shift to lower masses with increasing velocity dispersion predicted by the analytical models discussed in Section~\ref{sec:introth}.
We discuss a possible explanation for these results in the next section. 

\section{Discussion}
\label{sec:dis}
The purpose of our study is to test with numerical experiments the analytic theories presented in Section~\ref{sec:introth} and their predictions about the effect of turbulent fragmentation of the shape of the CMF/IMF. It is probably useful at this point to recall that \citet{padoan2002}, as well as \citet{hennebelle2008} and \citet{hopkins2012} assume an isothermal EOS at all densities. Besides, their results refer to the Core Mass Function (CMF), the mass distribution of the dense cores forming under the effect of turbulence and gravity, rather than to the IMF. The same predictions would hold also for a protostellar mass distribution if CMF and IMF are self-similar (see e.g.\ \citealt{alves2007}, \citealt{simpson2008}). 

We also recall that in the runs with lower mass resolution, despite the possibility of capturing low Jeans mass fragmentation at least partially ($m_{\rm res}\sim0.03M_{\odot}$), increasing the velocity dispersion of the gas seems to suppress the formation of low-mass objects. This results in a shift of the protostellar mass distribution towards higher masses, contrary to what is predicted by analytic theories. In the higher mass resolution simulations, on the other hand, we see no clear trend relating the protostellar mass function to the Mach number of the turbulence -- we neither see the shift to high masses found at low resolution nor the shift to low masses predicted by theory. We therefore have two important questions to answer: why do our low resolution and high resolution simulations behave differently? And why do our results fail to agree with the theoretical models?

The first of these questions is relatively simple to answer. As we have already mentioned, increasing the mass resolution of the simulation allows us to capture much more of the small-scale fragmentation that is missed in the low-resolution runs. Specifically, it allows us to follow the formation and fragmentation of protostellar discs that would have collapsed to a single sink particle in the low-resolution simulations, as shown in Figure~\ref{im_high}, thus populating the low-mass end of the IMF. 

Disc fragmentation also provides a simple answer to the question of why our results differ from the theoretical predictions in the case of our high resolution runs. All of the theoretical models discussed in Section~\ref{sec:introth} assume that turbulence plays the dominant role in generating the density structure of the gas and that the only forces resisting gravitational collapse are thermal pressure and turbulent pressure. However, these assumptions break down on the scale of individual protostellar accretion discs, where angular momentum plays a dominant role in structuring the gas and supporting it against collapse. It is therefore not surprising that the theoretical models do not properly represent the behaviour of the gas in this regime. 

\begin{figure}
	\includegraphics[scale=0.3]{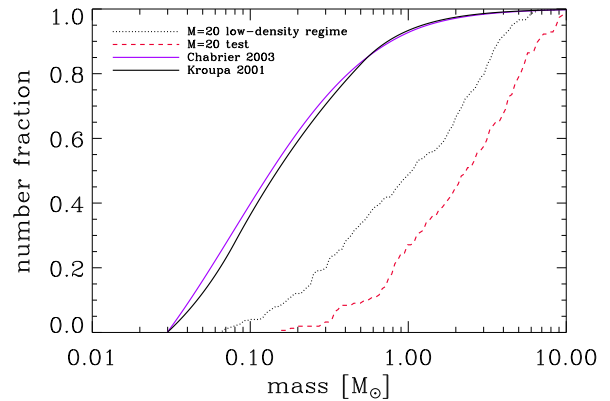}
	\caption{Comparison between the cumulative mass function in the low-density regime described in Section~\ref{low_res} for $\mathcal{M}=20$ and the cumulative mass function obtained for the same Mach number when shifting the knee of the EOS to $10^{-17} \text{g/cm}^3$ and thus stopping fragmentation at even lower densities.}
	\label{res_lowest}
\end{figure}

Additional proof of the importance of the EOS in shaping the system mass function is given by Figure~\ref{res_lowest}. It shows the results of a test simulation in which the initial conditions for the box (size, density, number of SPH particles) were chosen to be equal to those of the runs in the low-density regime, but the knee of the EOS was shifted to an even lower density of $10^{-17} \text{g/cm}^3$. We ran this test with a Mach number of $\mathcal{M}=20$ and compared the results with those from the simulation with identical Mach number in the low-density regime. The comparison confirms the conclusions derived above: the position of the knee of the EOS is crucial in determining the characteristic peak of the system mass function (see also \citealt{larson1985}, \citealt{larson2005}, \citealt{jappsen2005}).

\begin{figure}
	\includegraphics[scale=0.3]{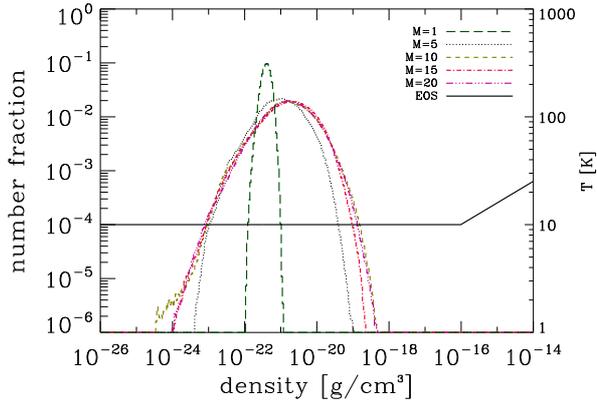}
	\caption{Mass-weighted density PDFs of the simulations in the low-density regime with different Mach numbers in the absence of gravity (broken lines). The solid line represents the EOS of the gas. The peak of the PDF lies at $\rho\sim10^{-21}\rmn{g \, cm^{-3}}$, far below the critical density for sink particle formation.}
	\label{mach_eos}
\end{figure}

There remains the question of why our simulations in the low-density regime find that the peak of the IMF shifts to higher masses with increasing turbulent velocity dispersion when the theoretical models predict the opposite. In this case, disc fragmentation does not operate in the simulations, and so we need to look for another way in which to explain the discrepancy.  In Figure~\ref{mach_eos} we plot the density PDF of the SPH particles for the runs with lower mass resolution and characterised by different Mach numbers together with the EOS described in Subsection~\ref{sec:eos}. The PDFs represent only the effects of turbulence on the gas, since they are taken from the last snapshot before we switch on self-gravity. All of the distributions peak at approximately the same density, $\rho\sim10^{-21}\rmn{g \, cm^{-3}}$, which is five orders of magnitude lower than $\rho_{\rm res}$
and seven orders of magnitude lower than the critical density for sink formation.

This means that in our simulations, turbulence alone cannot compress the gas up to densities that would immediately lead to sink formation. All of the work needed to make the gas collapse into protostellar objects is done by gravity and the properties of the dense cores produced by the turbulence therefore do not necessarily correlate well with the IMF that we observe at the end of the runs. In particular, the fact that turbulence tends to form denser cores, with lower Jeans masses, in runs with higher Mach numbers does not imply that these denser cores will form lower mass stars, since the difference in densities is small compared to the difference between these densities and $\rho_{\rm crit}$. 
Instead, a different effect appears to dominate as we increase the Mach number: it becomes increasingly easy to disrupt low-mass cores before they can collapse, and so the stars that form increasingly do so in high-mass cores (see also  \citealt{klessen2000} and \citealt{heitsch2001}, who make a similar argument).

\begin{figure}
 \includegraphics[scale=0.3]{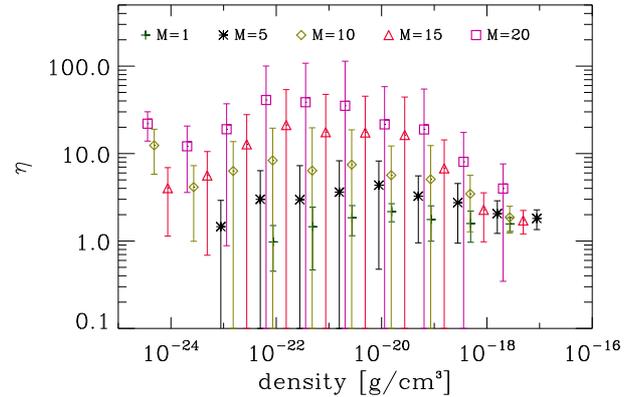}
 \includegraphics[scale=0.3]{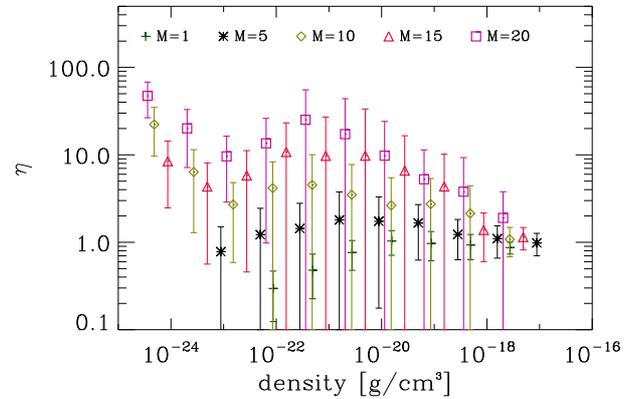}
  \caption{The virial parameter is plotted as a function of density. We chose several bins in density and randomly selected a sample of 100 SPH particles in each bin, for which we then calculated the average kinetic, thermal and potential energy with respect to the particles contained in one (top panel) and two (bottom panel) Jeans radii. This is done for all the simulations in the low-density regime. With increasing Mach number, the gas becomes less bound and the probability for collapse decreases.}
 \label{energy}
\end{figure}

Figure~\ref{energy} shows the virial parameter $\eta$:
\begin{equation}
 \eta=2(E_{\rm kin}+E_{\rm th})/E_{\rm pot},
\end{equation}
as a function of the density bin for different Mach numbers. In particular, we analyse the last snapshot before the creation of the first sink, since we are interested in the behaviour of the gas during gravitational collapse. 
In each density bin we randomly choose 100 test particles for which we calculate the thermal, potential and kinetic energy with respect to the gas enclosed in respectively one and two Jeans radii. In Figure~\ref{energy} the average over the 100 test particles for each bin is plotted. Despite the large scatter represented by the error-bars (enhanced by the logarithmic scale), it is evident that as we increase the velocity dispersion of the gas, dense cores become less and less bound. It is important to note that this is true not only for densities close to the mean cloud density, but also at the much higher densities corresponding to those found within the cores. 

As a consequence, it becomes increasingly difficult for protostellar objects (especially the less massive ones) to collapse further and form sink particles. While in the case of low Mach numbers ($\mathcal{M}=1-5$) the gas is bound almost over the entire range of densities due to the low kinetic energies, for higher Mach numbers, as expected, the gas is unbound at low densities and the ratio of kinetic+thermal and potential energy tends to decrease with density when the gas is about to collapse into a protostellar object. The gas being mostly unbound over the entire density range in simulations characterised by high Mach numbers renders the collapse of dense cores more difficult. 

Furthermore, shock waves travel faster in simulations with higher Mach number, and thus every point in space is hit by a shock front at smaller time intervals compared to simulations with lower velocity dispersion. The encounter between a dense core and a shock wave frequently leads to the disruption of the core and thus prevents the collapse and formation of protostellar objects. In particular the number of low-mass stars originating from small cores with low potential energies (that need a longer time to reach a density high enough for collapse) decreases, thus leading to a top-heavy IMF as we observe in Figure~\ref{res_driven}.

To summarize: increasing the Mach number of the turbulent flow leads to the mass distribution of the gravitationally-bound dense cores shifting to higher masses, simply because in these conditions, more massive cores are more likely to be gravitationally bound. In the low-density regime, fragmentation within these cores is suppressed and so the result is that the IMF also shifts towards higher masses. In the high-density regime, on the other hand, many of these cores fragment, reaching a state in which self-gravity dominates the local dynamical evolution of the fragments and causing the details of the initial conditions (in this case, the large-scale turbulent velocity field) to be rapidly forgotten (see, e.g., the discussions in \citealt{klessenburkert2000}, \citealt{klessen2001}, \citealt{maclowklessen2004}, \citealt{bate2005}, \citealt{clark2005}, \citealt{offner2014}).
In particular, disc fragmentation appears to become the dominant production mechanism for low-mass stars in this scenario, implying that theoretical models that do not account for its effects will be unable to produce the correct behaviour for the IMF. 

It should be stressed that the results from our high-density regime are very similar to those from other authors, who have employed different numerical codes/techniques. For example, our results are very similar to those of \citet{bate2009a}, who uses a higher resolution than we adopt here. Also \citet{delgadodonate2004} and \citet{girichidis2011} concluded that other factors than turbulence parameters play an important role in shaping the properties of the emerging sink particle mass functions. 

\section{Caveats}
\label{sec:cave}
Our simulations do not account for the radiative feedback of newly-formed stars on the surrounding gas and therefore may overestimate the amount of disc fragmentation that actually occurs. Whether or not this is the case is a matter of ongoing debate. 
\citet{bate2009a} has shown that simulations not including radiative feedback tend to overestimate the production of brown dwarfs. Including radiative feedback in subsequent simulations (\citealt{bate2009c}, \citealt{bate2012}), thus heating up the gas of the accretion discs, suppresses fragmentation and reproduces the observed IMF fairly well. Other simulations including radiative feedback alone, e.g. \citet{krumholz2012}, result in top-heavy mass distributions together with SFRs that are higher than observed. However, this result depends on the detailed numerical implementation of the process (e.g. \citealt{kuiper2011}), and is subject of ongoing debate (see also \citealt{peters2010a,peters2010b}). Including further physical processes, such as protostellar winds and driven turbulence in concert with radiative feedback  appears to improve the situation (see \citealt{krumholz2012}, \citealt{federrath2015}). Furthermore, \citet{stamatellos2012} claim that episodic accretion would allow the gas to cool down and to fragment between two bursts, while \citet{lomax2014} find that, although the absence of feedback leads to an overproduction of brown dwarfs, the continuous presence of feedback does not allow the formation of enough brown dwarfs to reproduce a Chabrier or Korupa-like IMF. They also show that a scenario characterised by episodic accretion seems to reproduce observations. 

We also note that fragmentation depends strongly on the thermodynamic properties of the gas, and hence on the effective equation of state that results from the balance between heating and cooling. In a cooling regime the gas reacts very strongly to pressure gradients. Turbulent compression can thus lead to large density contrasts which may become gravitationally unstable and collapse to form stars. If instead gas heats up when being compressed following $T\propto\rho^{\gamma-1}$ with $\gamma>4/3$, the critical mass for gravitational collapse increases and so the resulting stars become more massive. The amount of fragmentation occurring during gravitational collapse thus strongly depends on the compressibility of the gas \citep{li2003}. In real molecular clouds, the effective equation of state varies significantly as the gas density increases. At number densities  below $n \sim 10^5\,$cm$^{-3}$ the gas cools when being compressed by the emission of collisionally excited fine-structure or molecular lines (see e.g. \citealt{larson1985}, \citealt{larson2005}, \citealt{gloverclark2012a}). At higher densities the gas thermally couples to the dust and we expect that the temperature rises again with density \citep{banerjee2006}. This behavior sets a characteristic mass scale and defines the shape of the resulting IMF (see e.g. \citealt{jappsen2005}).

Other physical processes that are missing from our simulations include magnetic fields and protostellar feedback in the form of winds and outflows. These processes are expected to reduce the star formation rate within the clouds (see e.g.\ \citealt{klessen2000}, \citealt{heitsch2001},  \citealt{krumholz2010}, \citealt{wang2010}, \citealt{krumholz2011}, \citealt{federrath2015}) but their influence on the IMF is not yet fully understood (see e.g. \citealt{peters2011}).
We also note that other parameters such as the initial density profile and the ratio of compressive against solenoidal modes characterising the turbulent field can potentially influence the star formation process as well as the number of high- and low-mass stars forming, as shown e.g.\ in \citet{girichidis2011}, \citet{girichidis2012a,girichidis2012b},  and \citet{lomax2015}.

In addition, the results of our simulations show that the choice of the equation of state strongly influences the resulting mass distribution of the protostellar objects. In the case of the simulations in the low-density regime, the resolution density is $\sim3$ orders of magnitudes lower than the opacity limit, as discussed in Section~\ref{sec:eos}. Thus, the resulting mass distribution should be considered a system mass function resulting from the effect of turbulence and gravity on the gas, rather than an initial mass function. In the simulations in the high-density regime the resolution density is very similar to the opacity limit and thus reproduces more accurately, from this point of view, how fragmentation would proceed in nature. At densities lower than the resolution density we choose the EOS to be isothermal, not considering the cooling of collapsing gas at very low densities. Studying the effect of a varying EOS at low densities on the shape of the IMF should be investigated further in future studies.

\section{Conclusions}
\label{sec:concl}
We performed SPH simulations aimed at investigating the influence of supersonic turbulence on the mass spectrum of newly-formed stars. In the case of simulations in the low-density regime, in which fragmentation is artificially suppressed at relatively low densities (compared with the real opacity limit), we find that the position of the peak in the protostellar mass function and the underlying Mach number are correlated: an increase in the velocity dispersion of the star-forming gas corresponds to a shift of the characteristic mass of the distribution towards higher masses. 
This is due to the progressively increasing ratio between kinetic and potential energy that makes the collapse of low-mass cores less likely in a highly turbulent environment. The trend shown by our results is not consistent with the prediction of existing analytic theories investigating the influence of turbulence on the CMF. We argue that this is because in highly turbulent clouds, the dense cores formed by the turbulence are often not able to collapse due to the high kinetic and thermal energy of the gas. In addition, under these conditions it is more likely that dense cores are disrupted prior to collapse by shock waves travelling fast through the box.

In addition, we show that the resulting mass function of sink particles is very sensitive to the density at which we set the 'opacity limit' in our EOS (i.e. the transition from an effective barotropic gamma of 1 to 1.4). This demonstrates that in our low-density clouds, the EOS is as important as the turbulence at controlling the mass function of sinks. More importantly, the density at which we set the opacity limit is still an order of magnitude less than the densities reached during the standard turbulent compression, even in the most extreme case of a Mach number of 20.

In the high-density regime, the results do not show any trend when increasing the Mach number of the gas.  This implies that fragmentation on the scales of protostellar accretion discs is not affected by the large-scale dynamics of the turbulent field in the star-forming cloud.

In conclusion, our first set of simulations sample the low-density regime, in which turbulence is unable to shape the IMF in the same way as predicted by analytic theories, while the second set of simulations allow the gas to reach a higher-density regime in which turbulence does not seem to influence the shape of the IMF at all. 
	
It should be stressed however, that the simulations presented here are still unable to show full numerical convergence. Ideally, the results of our study could be tested by simulations that include both low- and high-density regimes at once and that allow the gas to reach the opacity limit, although this is currently beyond the state-of-the-art in computational resources. We argue that future analytic models for the IMF will have to address the physics of accretion discs and fragmentation more accurately in the future in order to correctly predict the shape of the IMF as a function of the environment.

\section*{Acknowledgments}
This work was supported by Sonderforschungsbereich SFB 881 "The Milky Way System" (sub-projects B1, B2, B5 and B8) of the German Research Foundation (DFG) and SPP 1573, ``Physics of the Interstellar Medium'' (grant number GL 668/2-1). In addition, RSK
acknowledges support from the European Research Council under the European Community's Seventh Framework Programme 
(FP7/2007-2013) via the ERC Advanced Grant STARLIGHT (project number 339177). The numerical simulations presented in this work were performed on the Milky Way supercomputer, which is funded by the Deutsche Forschungsgemeinschaft (DFG) through the Collaborative Research Center (SFB 881) "The Milky Way System" (subproject Z2) and hosted and co-funded by the J\"ulich Supercomputing Center (JSC). The authors thank the anonymous referee for helpful and constructive comments. CBM thanks Patrick Hennebelle, Pavel Kroupa, Tilman Hartwig and Scott Trager for helpful discussions.


\bibliographystyle{mnras}
\bibliography{references} 



\label{lastpage}
\end{document}